\documentclass[journal]{IEEEtran}
\ifCLASSINFOpdf
\else
\fi

\usepackage{stackengine}
\usepackage{multirow}
\usepackage{pgfplots}
\usepackage{tikzscale}
\newcommand\IncG[2][]{\addstackgap{%
		\raisebox{-.5\height}{\includegraphics[#1]{#2}}}}

\begin{document}
%
\title{Using Monte Carlo dropout for non-stationary noise reduction from speech.}
%
%

%
\author{Nazreen P.M. and A.G. Ramakrishnan,~\IEEEmembership{Senior Member, IEEE}
\thanks{Nazreen P.M. and A.G. Ramakrishnan  is with the Department
of Electrical Engineering, Indian Institute of Science, Bangalore,
India, 560012 e-mail: (nazreenp@iisc.ac.in and agr@iisc.ac.in).}}
\maketitle

\begin{abstract}
In this work, we propose the use of dropout as a Bayesian estimator for increasing the generalizability of a deep neural network (DNN) for speech enhancement. By using Monte Carlo (MC) dropout, we show that the DNN performs better enhancement in unseen noise and SNR conditions. The DNN is trained on speech corrupted with Factory2, M109, Babble, Leopard and Volvo noises at SNRs of 0, 5 and 10 dB.  Speech samples are obtained from the TIMIT database and noises from NOISEX-92. In another experiment, we train five DNN models separately on speech corrupted with Factory2, M109, Babble, Leopard and Volvo noises, at 0, 5 and 10 dB SNRs. The model precision (estimated using MC dropout) is used as a proxy for squared error to dynamically select the best of the DNN models based on their performance on each frame of test data. We propose an algorithm with a threshold on the model precision to switch between classifier based model selection scheme and model precision based selection scheme. Testing is done on speech corrupted with unseen noises White, Pink and Factory1 and all five seen noises.
\end{abstract}


%
\IEEEpeerreviewmaketitle

\section{Introduction}
Speech enhancement techniques find several applications such as automatic speech recognition, speaker recognition and hearing aids. Single channel speech enhancement has been a challenging problem for decades. 
 Several speech enhancement techniques have been proposed in the past. Methods such as spectral subtraction \cite{ss, GA}, Wiener filtering \cite{wiener}, minimum mean-square error (MMSE) estimators \cite{ephraim1984speech}, estimators based on Gaussian prior distributions \cite{martin2005speech, erkelens2007minimum} and residual-weighting schemes \cite{yegnanarayana1999speech, jin2006speech, prasanna2011} falls into the category of unsupervised enhancement methods. Most of these methods fail when the background noise is non-stationary and in unexpected acoustic conditions.

Supervised learning methods are expected to perform better than the unsupervised cases as prior information is being used \cite{srinivasan2006codebook, ephraim1992bayesian, sameti1998hmm}. To learn the complex mapping between noisy and clean speech, neural networks have been shown to be useful. Several models have been proposed in this field \cite{tamura1989analysis, xie1994family, wan1999networks}. However these models are small to properly learn the complex mapping. Deep architectures have been widely used in this area recently as they have shown the ability to learn the complex mapping between noisy and clean features and hence give superior enhancement performances. Hinton \textit{et al.} proposed a greedy layer-wise unsupervised learning algorithm \cite{hinton2006reducing, hinton2006fast}.
Mass et al. \cite{maas2012recurrent} use deep recurrent neural networks (DRNNs) for feature enhancement for noise robust ASRs.  

One of the major issues encountered by DNN based enhancement is the degradation of performance on  noises for which the network is not adapted which is referred to as unseen condition. The model learns mapping between noisy and clean speech well for noises and SNRs on which it is trained, but performs poor on speech corrupted by an unseen noise or SNR. In fact, this itself could be dealt with as a challenging task in speech enhancement scenario. Though not dealt with separately, several techniques have been proposed in the past to address this problem. \cite{xu2014experimental} proposed a regression DNN-based speech enhancement framework, where a wide neural network is trained using a large collection of data of about 100 hours of various noise types. A DNN-SVM based system is proposed in \cite{wang2013towards}, which is trained on a variety of acoustic data for a considerably huge amount of time. A noise aware training technique is adopted in \cite{xu2015regression}, where a noise estimate is also appended along with the input feature for training. They use about 2500 hours of training data for training the network. 

We propose a new algorithm to improve the performance of unseen noises using a new algorithm based on Monte Carlo dropout proposed by Gal and Ghahramani in \cite{gal2016dropout}. Our experiments show that the algorithm gives superior performance in most of the unseen noise cases compared to that using conventional dropout \cite{dropout, srivastava2014dropout}.

\section{Related Works}
Hinton \textit{et al.} \cite{dropout, srivastava2014dropout} introduced the concept of dropout to reduce overfitting during DNN training. Though dropout omits weights during training, it is inactive during the inference stage, whereby all the neurons contribute to the prediction.

Gal and Ghahramani in \cite{gal2016dropout} shows a theoretical relationship between dropout \cite{srivastava2014dropout}  and approximate inference in a
Gaussian process and introduced the method of using dropout during inference. Kendall \textit{et al.} in \cite{kendall2016modelling} shows that by enabling dropout during inference and averaging the results of multiple stochastic forward passes, the predictions show improvement. \cite{kendall2016modelling} uses the term MC droput to refer to this technique. These samples could be considered as Monte Carlo samples, from the posterior posterior distribution of models \cite{kendall2016modelling}. Gal \textit{et al.} \cite{gal2016dropout} also show the estimation of model uncertainty from these samples.

The focus of this work is to use the idea od MC dropout to improve the generalizability of speech models and to improve the enhancement performance in a highly mismatched condition. In \cite{mc_my1} we show that in the case of noisy speech corrupted with unseen noises, MC dropout models can give a better denoised output than conventional dropout models. To show this we train two DNN models on multiple noises and SNRs , one employing MC dropout and another employing the conventional dropout and compare the performance of the two.


We also explore the usage of model uncertainty in problems where multiple noise specific DNN models are used. By using model uncertainty as an estimate of the prediction error for a sample, this technique can enable selection of best models with least prediction error in frame basis. A similar approach of selecting the best model based on an error estimate was proposed in \cite{papadopoulos2016long}. However this was used for robust SNR estimation. They trained a separate DNN as a classifier to select a particular regression model for SNR estimation. However this approach does not ameliorate the original problem of mismatch in training and testing conditions. In our proposed algorithm we use the intrinsic uncertainty of a model to estimate the prediction error. Since this method extracts information from the model itself, it has the potential to be a better representative of the prediction error. Our method also circumvents the issue of unseen testing conditions since according to \cite{gal2016dropout}, the model uncertainty itself is an indicator of unseen data. This paper extends our preliminary results reported in \cite{mc_my1}. We propose a predictive variance threshold based algorithm to switch between model uncertainty based selection scheme and classifier based model selection scheme to compensate for the performance drop of the intrinsic uncertainty based algorithm for seen noises.


The baseline system is augmented by MC dropout as a bayesian approximation. The distribution over the weights could be learned using this approximation consequently giving an uncertainty of the output.
%
The input $X$ is fed into the network using dropout same as that employed during the training time. Multiple passes are made through the network dropping out different random units each time. Thus $T$ repetitions are performed by dropping of random units each time during testing. This results in $ T$ different outputs for a given input $ X$ ; $\{\hat{S_{t}(X)}\};1\leq t\leq T$. \cite{gal2016dropout} shows that averaging forward passes through the network is equivalent to Monte Carlo integration over a Gaussian process posterior approximation. Empirical estimators of the predictive mean ($E(S)$) and variance (uncertainty, $Var(S)$) from these samples are given as:

\begin{equation}
E(S) \approx \frac{1}{T}\sum_{t=1}^{T} \hat{S_{t}(X)}
\label{eq5}
\end{equation}
\begin{equation}
Var(S)\approx \tau ^{-1} I_D + \frac{1}{T}\sum_{t=1}^{T} \hat{S_{t}(X)}^T \hat{S_{t}(X)} - E(S)^T E(S)
\label{eq6}
\end{equation}
where $\tau = {l^2 p}/{2N \lambda}$ ; $l$: defined prior length scale, $p$: probability of the units not being dropped, $N$: total input samples,  $\lambda$: regularisation weight decay, which is zero for our experiments.

\section{DNN based speech enhancement}
Under additive model, the noisy speech can be represented as,
\begin{equation}
	x_t(m)=s_t(m)+n_t(m)
\end{equation}
where $ x_t(m) $, $ s_t(m) $ and $ n_t(m) $ are the $m^{th}$ samples of the noisy speech, clean speech and noise signal, respectively. Taking the short time Fourier transform (STFT), we have,
\begin{equation}
	x(\omega_k)=s(\omega_k)+n(\omega_k)
\end{equation}
where $\omega_k = (2 \pi k/R)$, $k=0,1,2...R-1$, $k$ is the index and $R$ is the number of frequency bins. Taking the magnitude of the STFT, the noisy speech can be approximated as
\begin{equation}
\end{equation}
where $ S $ and $N$ represent the spectra of the clean speech and the noise, respectively. 

A DNN based regression model is trained using the magnitude STFT features of clean and noisy speech. The noisy features are then fed to this trained DNN to predict the enhanced features, $\hat{S}$. The enhanced speech signal is obtained by using the inverse Fourier transform of $\hat{S}$ with the phase of the noisy speech signal and overlap-add method.

\subsection{Baseline DNN architecture} \label{archi}
The baseline DNN consists of 3 fully connected layers of 2048 neurons and an output layer of 257. We use ReLU non-linearity as the activation function in all the 3 layers. Our output activation is also ReLU to account for the nonnegative nature of STFT magnitude. 
Stochastic gradient descent is used to minimize the mean square logarithmic error ($E_{r}$) between the noisy and clean magnitude spectra:
\begin{equation}
	E_{r} =\frac{1}{R} \sum_{k=1}^{R}(log(S(k)+1)-log(\hat{S(k)}+1))^2
\end{equation}
where $\hat{S}$ and $S$ denote the estimated and reference spectral features, respectively, at sample index $k$. 



\section{Proposed methods for generalized speech models}

Our approach to improve generalisation involves two approaches. In the first approach, we show that MC dropout estimate shows improvement in the generalization performance of DNN and apply this to speech enhancement.

 In the second approach we use model uncertainty to optimally choose among multiple DNN models so that the reconstruction error is minimum. 
 This analysis involves two sets of frameworks as explained in sec. \ref{multi1} and  \ref{multi2}.


\subsection{Single DNN model using MC dropout (single-MC)}\label{single}

In this method we use MC dropout to improve the genarilzability of the baseline model. To evaluate the proposed method, we train a DNN model using MC dropout and evaluate the performance against the one using conventional dropout. 

\begin{figure}[ht!]%
	
	
	\centering

	\includegraphics[height=2.2cm,width=8.5cm]{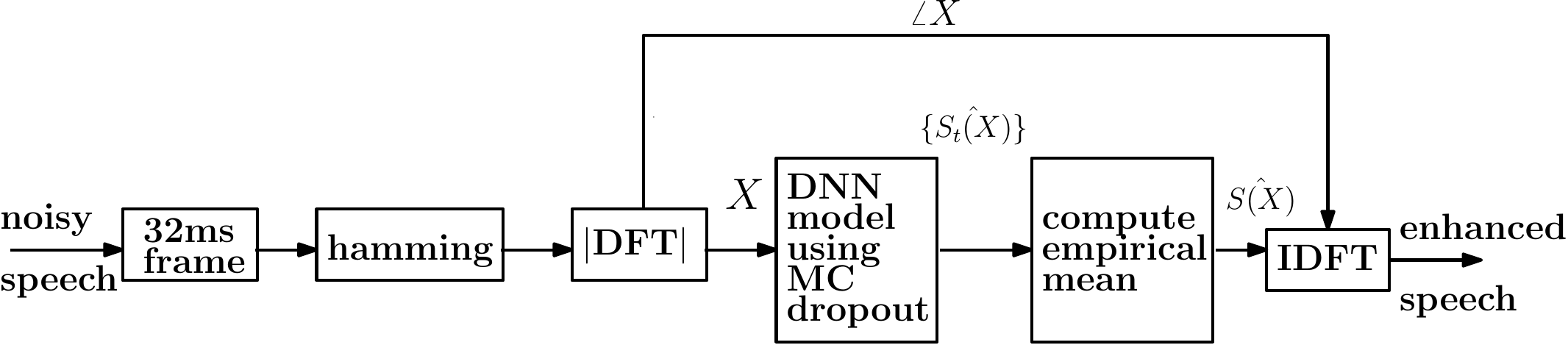}

	
	\caption{Enhancement using single DNN-MC dropout model.}
	\label{fig1}	
\end{figure}

A single DNN model is trained using speech signals corrupted with various noises and SNRs employing MC dropout. The block diagram of the proposed approach is shown in Figure \ref{fig1}. The input noisy speech is divided into frames and STFT is applied. Let $ X$ denote the magnitude STFT feature for a particular frame. Given a noisy speech frame $X$, multiple repetitions are performed by dropping out random units each time giving $T$ different outputs, $ \{\hat{S_{t}(X)}\}; 1\leq t\leq T$. The empirical mean of these outputs \ref{eq5} is  the estimated output $\hat{S(X)}$. Enhanced speech is obtained as the inverse Fourier transform of $\hat{S(X)}$ with the phase of the noisy speech signal and overlap-add method.
\begin{equation}
	\hat{S(X)} \approx \frac{1}{T}\sum_{t=1}^{T} \hat{S_{t}(X)}
	\label{eq7}
\end{equation}
\begin{equation}
	\hat{s(x)} = IDFT(\hat{S(X)} \angle X)
	\label{eq8}
\end{equation}
where $\hat{s(x)}$ indicates the enhanced speech estimate for a noisy speech input $x$
\subsection{Multiple noise-specific MC dropout models for enhancement}
Model-specific enhancement techniques depend on a model selector, which ensures that the model chosen for enhancing each frame entails an overall improved performance \cite{dnn_phn, my}. Given a framework of multiple DNN models for enhancement, one needs to select the appropriate noisy model to enhance the input noisy speech frame. One of the methods that one could employ is to use a noise classifier \cite{papadopoulos2016long} to select the appropriate noise model. However in the scenarios where the input speech is corrupted with an unknown noise or SNR condition, the noise classifier might fail to pick the optimal model. In these cases, we need to ensure that the chosen model is the one that gives the lowest error and hence a better enhancement performance. In our methods \ref{multi1} \ref{multi2}, we use the model uncertainty estimated from the output samples of each MC dropout model as an estimate of the prediction error and choose the model based on it. Our experiments show that stronger the correlation between model uncertainty and the squared error, better is the enhancement performance. For evaluating the performance, we compare our algorithms with the one where a classifier is used to pick the noise model. Here the noise model could be one using MC dropout (class-MC) and using conventional dropout (class-Conv) 

\subsubsection{Multiple models using MC dropout with predictive variance (model uncertainty) as the selection scheme (Var-MC)}\label{multi1}
\begin{figure}[ht!]%
	
	
	\centering

	\includegraphics[height=5cm,width=8.8cm]{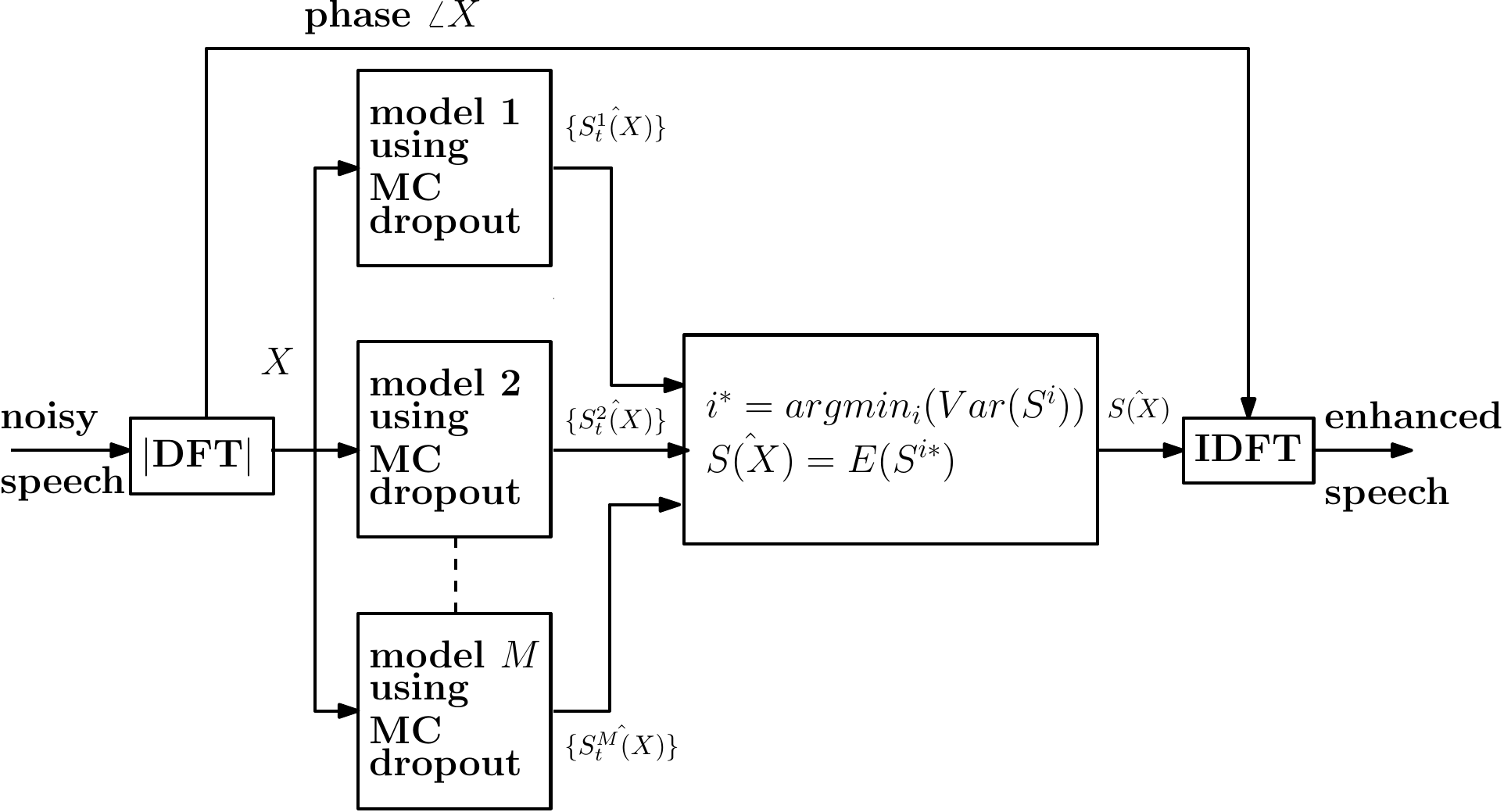}


	\caption{Enhancement using multiple DNN-MC dropout models with predictive variance as the selection criteria.}
		\label{fig2}	
\end{figure}


In this work, we follow \cite{gal2016dropout} and say that since model uncertainty gives the intrinsic uncertainty of the model for a particular input, we can use it as an estimate of model error. The speech enhancement framework so designed is as shown in fig. \ref{fig2}.


$M$ different DNN models with MC dropout are trained with speech corrupted with various noises and SNRs. The architecture of each model is as mentioned in section \ref{archi}. Input noisy speech is first divided into frames and magnitude STFT is obtained. The input noisy magnitude STFT feature $X$ of a frame is fed into each of these five models.  $T$ repetitions are performed by each model by dropping different units every time, obtaining results $ \{\hat{S^i_{t}(X)}\}; 1\leq t\leq T ; 1\leq i\leq M$ ; where $i$ is the model index. The predictive variance (model uncertainty) of each of these $M$ outputs are computed . The output with the minimum variance, $ \{\hat{S^{i*}_{t}(X)}\}; 1\leq t\leq T ; 1\leq {i*}\leq M$ ; is selected and the corresponding model is considered the best for that particular input $X$. The enhanced output $\hat{S}$ is estimated as the empirical mean of the $T$ outputs: $ \{\hat{S^{i*}_{t}(X)}\}; 1\leq t\leq T$. The enhanced speech signal is obtained as the inverse Fourier transform of $\hat{S}$ with the phase of the noisy speech signal and overlap-add method.

\begin{equation}
	\hat{S(X)} \approx \frac{1}{T}\sum_{t=1}^{T} \hat{S^{i*}_{t}(X)}
	\label{eq9}
\end{equation}

  \subsubsection{A predictive variance threshold ($\mu$) based algorithm for enhancement using multiple models ($\mu$-MC)}\label{multi2}
  The experimental results of algorithm Var-MC shows superior performance for most of the unseen noises. However, the performance on seen noise shows significant degradation. This can be rectified using a conditional selection criteria for the noise models. Using this condition, selection of  noise models can be switched from model uncertainty based to classifier based.

\begin{figure}[ht!]%
  	
  	
  	\centering

  	\includegraphics[height=5cm]{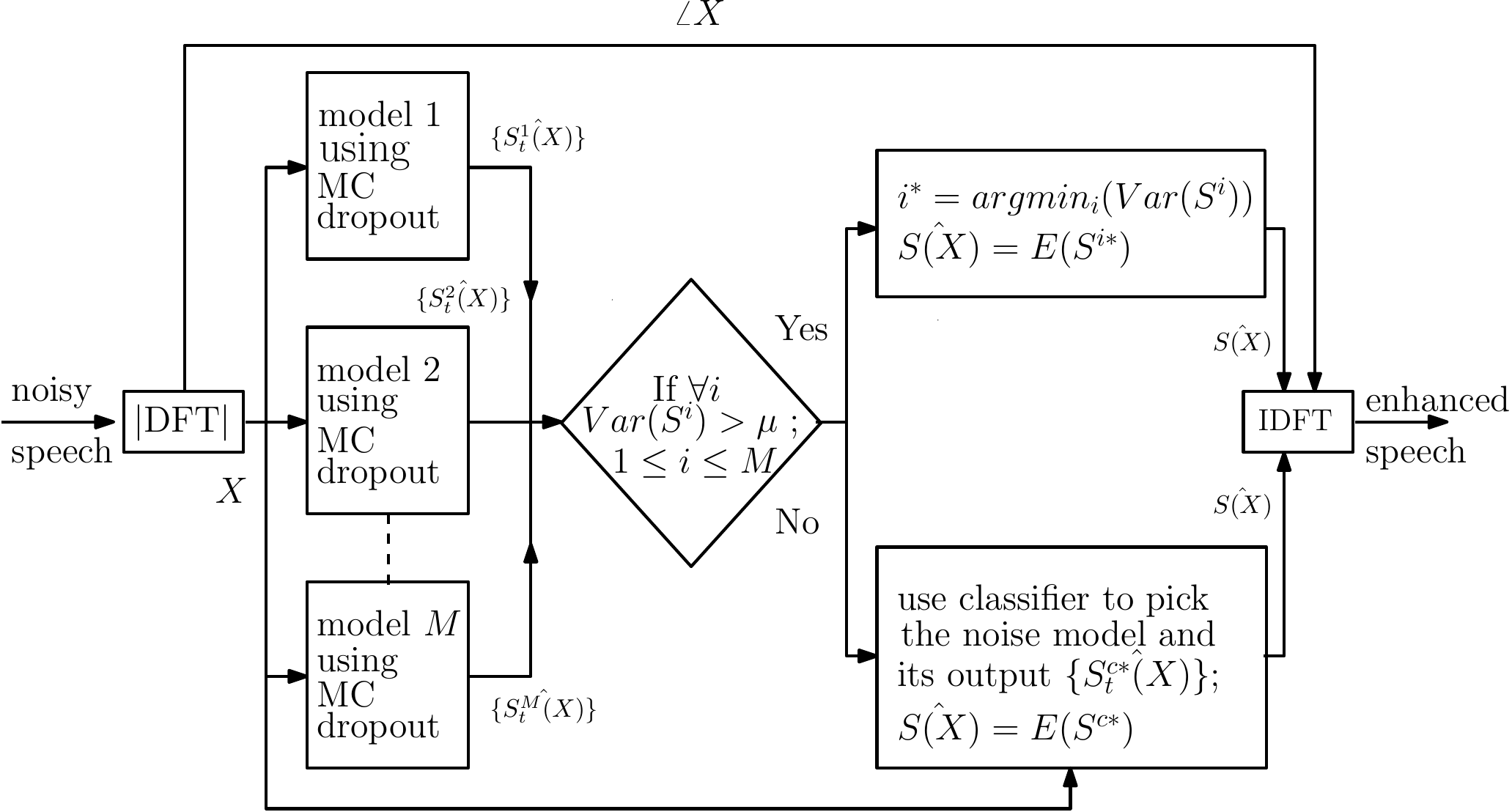}

  	
  	\caption{A predictive variance threshold ($\mu$) based algorithm for enhancement using multiple models}
  	\label{fig3}	
  \end{figure}

A threshold is set for variance of all the five models, so that the model for enhancing a noisy  frame could either be selected on the basis of minimum variance scheme or on the basis of the prediction of a noise classifier as shown in Figure \ref{fig3}.

The input noisy feature of a frame $X$, is fed into all the five MC dropout models. The input is passed $T$ different times by dropping out random units each time.  The corresponding outputs are $ \{\hat{S^i_{t}(X)}\}; 1\leq t\leq T ; 1\leq i\leq M$ ; where $i$ is the model index and $M=5$. Then the predictive variance  $ V(S^i)$ of each of these $M$ outputs are computed. If all the $M$ uncertainty values are above a threshold, say $\mu$, it could be taken as an indication that the noise corrupting the given input speech belongs to none of these $M$ noise models. In such a case, the model which gives the minimum value of uncertainty is considered as the best model to enhance the input noisy speech feature $X$. the corresponding output is; 
$ \{\hat{S^{i*}_{t}(X)}\}; 1\leq t\leq T ; 1\leq {i*}\leq M$. Taking the empirical mean of these $T$ output gives the enhanced output.
\begin{equation}
	\hat{S(X)} \approx \frac{1}{T}\sum_{t=1}^{T} \hat{S^{i*}_{t}(X)}
	\label{eq10}
\end{equation}
The enhanced speech signal is obtained as the inverse Fourier transform of $\hat{S}$ with the phase of the noisy speech signal and overlap-add method.

 On the other hand if the uncertainty values are below the threshold $\mu$, the input feature $X$ is first fed into a classifier to decide the best model for enhancing the frame. Let the corresponding output be;  $ \{\hat{S^{c*}_{t}(X)}\}; 1\leq t\leq T ; 1\leq {c*}\leq M$. As mentioned previously, taking the empirical mean of these $T$ different outputs gives the enhanced output $\hat{S}$.
     \begin{equation}
     	\hat{S(X)} \approx \frac{1}{T}\sum_{t=1}^{T} \hat{S^{c*}_{t}(X)}
     	\label{eq11}
     \end{equation}
The enhanced speech is obtained as the inverse Fourier transform with the noisy phase information and overlap add method.     

  \section{Experimental setup}
  
  All experiments are carried out using TIMIT \cite{timit} speech corpus. The noise data is obtained from NOISEX-92 \cite{noisex} database. In-order to synthesize noisy test and training speech data, the noise files are downsampled to 16 kHz so as to match with the sampling rate of TIMIT. The magnitude STFT is computed on frames of size 32 ms with 10 ms frame shift, after applying Hamming window. A 512-point FFT is taken and we use only the first 257 points as input to the DNN, because of symmetry in the spectrum. 
  
  For our experiments, the number of repetitions $T$ is chosen as 50. The Adam optimizer \cite{kingma2014adam} is chosen, whose default regularization weight decay, $\lambda$ is zero and thus, $\tau^{-1} = 0$ in eqn.\ref{eq6}.

  \subsection{single-MC}
  Each DNN based regression models are trained with the magnitude STFT of noisy speech as input and clean speech as target. For experiments using single DNNs \ref{single}, a baseline DNN with conventional dropout \cite{dropout, srivastava2014dropout}and a DNN using MC dropout is trained using speech corrupted with factory 2, m109, leopard, babble and volvo noises at 0, 5 and 10 dB SNRs. The architecture of both models are as mentioned in section \ref{archi}. The training is done on the entire TIMIT training data after randomly dividing them into fifteen parts for adding five noises at three different SNRs.
  \subsection{Var-MC and $\mu$-MC}
  For multiple DNN model based experiments \ref{multi1} \ref{multi2}, five DNN models are trained on speeches corrupted with factory2, m109, leopard, babble and volvo noises, each at SNRs 0, 5 and 10 dB. Each DNN models are trained using MC and conventional dropout, using the entire TIMIT training data after randomly dividing the files into three for adding noises at SNRs 0, 5 and 10 dB. In this case also, the architecture of the models are as defined in section \ref{archi}.     
  
 For those experiments where a classifier is used to pick the models (class-MC and class-Conv) , the classifier used is trained on speech corrupted with factory2, babble, leopard, m109 and volvo noises at SNRs 0,5 and 10dB.

  \section{Results and discussion}
    
  \subsection{single-MC}
   \begin{table*}[]
      	\centering
      	\caption{Performance evaluation of single DNN model with MC dropout (single-MC)}
      	
      	
      	\resizebox{0.99\linewidth}{2.9cm}{
      		\begin{tabular}{|c|c|c|c|c|c|c|c|c|c|c|c|c|c|}
      			\hline
      			\textbf{}                     & \textbf{}     & \multicolumn{3}{c|}{\textbf{White (unseen)}}                                                                          & \multicolumn{3}{c|}{\textbf{Pink (unseen)}}                                                                           & \multicolumn{3}{c|}{\textbf{Factory1 (unseen)}}                                                                      & \multicolumn{3}{c|}{\textbf{Factory2 (seen)}}                                 \\ \hline
      			\textbf{SNR}                  & \textbf{Metric}     & \textbf{\begin{tabular}[c]{@{}l@{}}Noisy\\input\end{tabular}}                     & \textbf{single-conv}                  & \textbf{single-MC}                        & \textbf{\begin{tabular}[c]{@{}l@{}}Noisy\\input\end{tabular}}                     & \textbf{single-conv}                  & \textbf{single-MC}                        & \textbf{\begin{tabular}[c]{@{}l@{}}Noisy\\input\end{tabular}}                     & \textbf{single-conv}                  & \textbf{single-MC}                       & \textbf{\begin{tabular}[c]{@{}l@{}}Noisy\\input\end{tabular}}                     & \textbf{single-conv} & \textbf{single-MC}   \\ \hline
      			\multirow{2}{*}{\textbf{-10}} & \textbf{SSE x10\textasciicircum 4}  & {3.64} & {3.36} & \textbf{3.14} & {3.96} & {0.874} & \textbf{0.848} & {3.69} & {0.720} & \textbf{0.70} & {4.13} & {0.0467}      & \textbf{0.0461}  \\ \cline{2-14} 
      			& \textbf{SSNR} & {-8.9}                      & {-8.5}                      & \textbf{-8.4}                      & {-8.8}                      & {-6.7}                      & \textbf{-6.6}                      & {-8.7}                      & {-6.0}                      & \textbf{-5.9}                     & {-8.5}                      & {1.0}      & {1.0}  \\ \hline
      			
      			\multirow{2}{*}{\textbf{-5}} & \textbf{SSE x10\textasciicircum 4}  & {1.12} & {0.960} & \textbf{0.913} & {1.22} & {0.270} & \textbf{0.251} & {1.12} & {0.213} & \textbf{0.200} & {1.29} & {0.0198}      & \textbf{0.0197}  \\ \cline{2-14} 
      			& \textbf{SSNR} & {-7.2}                      & {-6.6}                      & \textbf{-6.5}                      & {-7.1}                      & {-4.3}                      & \textbf{-4.2}                      & {-6.9}                      & {-3.51}                     & \textbf{-3.50}                     & {-6.7}                      & {3.05}     & {3.08} \\ \hline
      			\multirow{2}{*}{\textbf{0}} & \textbf{SSE x10\textasciicircum 3}  & {3.41} & {2.81 } & \textbf{2.60} & {3.71} & {0.858}      & \textbf{0.843}  & {3.41} & {0.682}      & \textbf{0.671}   & {4.01} &{0.104}      & {0.104}  \\ \cline{2-14} 
      			& \textbf{SSNR} & {-4.6}                      & {-3.9}                     & \textbf{-3.8}                      & {-4.5}                      & {-1.5}     & \textbf{-1.4} & {-4.4}                      & {-0.73}    & {-0.73} & {-4.1}                      & {5.1}      &{5.1}  \\ \hline
      			
      			\multirow{2}{*}{\textbf{5}} & \textbf{SSE x10\textasciicircum 3}  & 1.03 & 0.844               & \textbf{0.827}         & 1.12 & 0.291               & \textbf{0.288}         & 1.02 & 0.244               & \textbf{0.242}         & 1.24 & 0.069                & 0.069          \\ \cline{2-14} 
      			& \textbf{SSNR} & -1.6                      & -0.7              & -0.7        & -1.4                      & 1.7               & 1.7         & -1.3                      & 2.2               & 2.2         & -0.9                      & 7.1               & 7.1         \\ \hline
      			\multirow{2}{*}{\textbf{10}} & \textbf{SSE x10\textasciicircum 2}  & 3.08            & 2.70               & \textbf{2.67}         & 3.41            & 1.18               & \textbf{1.16}         & 3.09            & 1.07               & \textbf{1.06}         & 3.82            & 0.56                & 0.55          \\ \cline{2-14} 
      			& \textbf{SSNR} & 2.0            & 2.7               & 2.7         & 2.2            & 4.7               & 4.7         & 2.3            & 5.0               & 5.0         & 2.6            & 8.9               & 8.9         \\ \hline
      			
      		\end{tabular}
      	}
      	
      	\label{table1}
      \end{table*}

  Table \ref{table1} shows the results obtained in terms of sum squared error (SSE), and segmental SNR (SSNR) \cite{hu2008evaluation} for single DNN-MC dropout model (single-MC) over the baseline (single-conv) for unseen and seen noises. We use white, pink and factory 1 noise as unseen noises and factory2 as a seen noise. The reported results are the average over 50 files randomly selected from TIMIT \cite{timit} test set. From the table it can be inferred that MC dropout model achieves superior performance in most of the  unseen noise cases. It is to be noted that the improvement is significant for unseen noises like white noise, especially at low SNRs of -10 and -5 dB. Interestingly, the performance degrades at higher SNRs, though the model continues to perform better than the baseline (single-conv) in terms of SSE. Though the proposed method does not result in significant improvement on seen noises, the performance is comparable to the baseline model. Hence, the observations validate the proposed method of using MC dropout to improve generalization performance on unseen noises.

\subsection{Var-MC}

Tables \ref{table2} and \ref{table3} shows the  performance of our Var-MC algorithm in terms of SSE and SSNR. It can be inferred from the table that, Var-MC gives superior performance over class-conv and class-MC for most of the unseen noise cases.  However as the SNR improves, the improvement over the baseline drops. This performance drop can be explained by the reduced correlation between the squared error and the model uncertainty that is shown in Fig. \ref{fig4}. 
   \begin{table*}[]
                 	\centering
                 	\caption{Performance evaluation of Var-MC and $\mu$-MC algorithms.}
                 	
       
                 	\resizebox{0.99\linewidth}{3.2cm}{
                 		\begin{tabular}{|c|c|c|c|c|c|c|c|c|c|c|c|c|c|c|c|c|c|c|c|c|c|}
                 			\hline
                 			&                                     & \multicolumn{5}{c|}{\textbf{White (unseen)}}                                                                                                                       & \multicolumn{5}{c|}{\textbf{Pink (unseen)}}                                                                                                                        & \multicolumn{5}{c|}{\textbf{Factory1 (unseen)}}                                                                                                                    & \multicolumn{5}{c|}{\textbf{Factory2 (seen)}}                                                                                                                    \\ \hline
                 			\textbf{SNR}                     & \textbf{Metric}                     & \textbf{\begin{tabular}[c]{@{}l@{}}Noisy\\input\end{tabular}} & \textbf{Class-conv} & \textbf{Class-MC} & \textbf{Var-MC} & \textbf{\begin{tabular}[c]{@{}l@{}}$\mu$-MC\\$\mu=0.16$\end{tabular}} & \textbf{\begin{tabular}[c]{@{}l@{}}Noisy\\input\end{tabular}} & \textbf{Class-conv} & \textbf{Class-MC} & \textbf{Var-MC} & \textbf{\begin{tabular}[c]{@{}l@{}}$\mu$-MC\\$\mu=0.16$ \end{tabular}} & \textbf{\begin{tabular}[c]{@{}l@{}}Noisy\\input\end{tabular}} & \textbf{Class-conv} & \textbf{Class-MC} & \textbf{Var-MC} & \textbf{\begin{tabular}[c]{@{}l@{}}$\mu$-MC\\$\mu=0.16$\end{tabular}} & \textbf{\begin{tabular}[c]{@{}l@{}}Noisy\\input\end{tabular}} & \textbf{Class-conv} & \textbf{Class-MC} & \textbf{Var-MC} & \textbf{\begin{tabular}[c]{@{}l@{}}$\mu$-MC\\$\mu=0.16$\end{tabular}} \\ \hline
                 			\multirow{2}{*}{\textbf{-10 dB}} & \textbf{SSE x10\textasciicircum{}4} & 3.64           & 3.61              & 3.42             & \textbf{3.23}  & \textbf{3.24}                                                                    & 3.96           & 1.13              & 1.17             & \textbf{0.708} & \textbf{1.05}                                                                    & 3.69           & 1.03              & 1.01             & \textbf{0.677} & \textbf{0.876}                                                                   & 4.13           & 0.0406            & 0.0397           & 0.331            & 0.0458                                                                           \\ \cline{2-22} 
                 			& \textbf{SSNR}                       & -8.9           & -8.7              & -8.6             & \textbf{-8.4}  & \textbf{-8.5}                                                                    & -8.8           & -7.1              & -7.1             & \textbf{-5.4}  & \textbf{-6.9}                                                                    & 8.7            & -6.6              & -6.6             & \textbf{-5.3}  & \textbf{-6.3}                                                                    & -8.5           & 2.1               & 2.1              & 0.5            & 2.1                                                                              \\ \hline
                 			\multirow{2}{*}{\textbf{-5 dB}}  & \textbf{SSE x10\textasciicircum{}4} & 1.12           & 1.02              & 0.976            & \textbf{0.936} & \textbf{0.956}                                                                   & 1.22           & 0.312             & 0.322            & \textbf{0.261} & \textbf{0.311}                                                                   & 1.12           & 0.285             & 0.285            & \textbf{0.20}  & \textbf{0.260}                                                                   & 1.29           & 0.0172            & 0.0171           & 0.257            & 0.0259                                                                           \\ \cline{2-22} 
                 			& \textbf{SSNR}                       & -7.2           & -6.7              & -6.6             & \textbf{-6.5}  & -6.6                                                                             & -7.1           & -4.5              & -4.5             & \textbf{-3.7}  & -4.5                                                                             & -6.9           & -4.1              & -4.1             & \textbf{-3.3}  & \textbf{-4.0}                                                                    & -6.7           & 4.0               & 4.0              & 1.3            & 3.9                                                                              \\ \hline
                 			\multirow{2}{*}{\textbf{0 dB}}   & \textbf{SSE x10\textasciicircum{}3} & 3.41           & 2.94              & 2.86             & \textbf{2.70}  & \textbf{2.84}                                                                    & 3.71           & 0.902             & 0.918            & 0.943          & 0.981                                                                            & 3.41           & 0.828             & 0.832            & \textbf{0.771} & 0.836                                                                            & 4.01           & 0.089             & 0.090            & 1.37            & 0.15                                                                             \\ \cline{2-22} 
                 			& \textbf{SSNR}                       & -4.6           & -4.1              & -4.0             & \textbf{-3.8}  & -4.0                                                                             & -4.5           & -1.6              & -1.6             & \textbf{-1.3}  & -1.6                                                                             & -4.4           & -1.1              & -1.1             & \textbf{-0.83} & -1.1                                                                             & -4.1           & 5.8               & 5.8              & 3.3            & 5.8                                                                              \\ \hline
                 			\multirow{2}{*}{\textbf{5 dB}}   & \textbf{SSE x10\textasciicircum{}3} & 1.03           & 0.884             & 0.865            & \textbf{0.857} & \textbf{0.856}                                                                   & 1.12           & 0.288             & 0.290            & 0.391          & 0.339                                                                            & 1.02           & 0.270             & 0.273            & 0.285          & 0.288                                                                            & 1.24           & 0.059             & 0.060            & 0.456           & 0.09                                                                             \\ \cline{2-22} 
                 			& \textbf{SSNR}                       & -1.6           & -0.8              & -0.8             & \textbf{-0.7}  & \textbf{-0.7}                                                                    & -1.4           & 1.7               & 1.7              & 1.6            & 1.7                                                                              & -1.3           & 2.0               & 2.0              & 2.0            & 2.0                                                                              & -0.9           & 7.7               & 7.7              & 5.8            & 7.6                                                                              \\ \hline
                 			\multirow{2}{*}{\textbf{10 dB}}  & \textbf{SSE x10\textasciicircum{}2} & 3.08           & 2.82              & 2.81             & \textbf{2.73}  & \textbf{2.69}                                                                    & 3.41           & 1.12              & 1.14             & 1.40           & 1.20                                                                             & 3.09           & 1.10              & 1.14             & 1.24           & 1.16                                                                             & 3.82           & 0.47              & 0.48             & 1.34           & 0.55                                                                             \\ \cline{2-22} 
                 			& \textbf{SSNR}                       & 2.0            & 2.6               & 2.6              & \textbf{2.7}   & \textbf{2.7}                                                                     & 2.2            & 4.8               & 4.8              & 4.5            & 4.7                                                                              & 2.3            & 4.9               & 4.9              & 4.8            & 4.9                                                                              & 2.6            & 9.5               & 9.5              & 8.1            & 9.5                                                                              \\ \hline
                 		\end{tabular}
                 	}
                 \label{table2}
                 \end{table*}

         \begin{table*}[]
                  	\centering
                  	\caption{Performance evaluation of Var-MC and $\mu$-MC algorithms.}
       
                  	  	\resizebox{0.99\linewidth}{3.2cm}{
                  	\begin{tabular}{|c|c|c|c|c|c|c|c|c|c|c|c|c|c|c|c|c|c|c|c|c|c|}
                  		\hline
                  		&                                     & \multicolumn{5}{c|}{\textbf{M109 (seen)}}                                                                                                                        & \multicolumn{5}{c|}{\textbf{Leopard (seen)}}                                                                                                                     & \multicolumn{5}{c|}{\textbf{Babble (seen)}}                                                                                                                      & \multicolumn{5}{c|}{\textbf{Volvo (seen)}}                                                                                                                       \\ \hline
                  		\textbf{SNR}                     & \textbf{Metric}                     & \textbf{\begin{tabular}[c]{@{}l@{}}Noisy\\input\end{tabular}} & \textbf{Class-conv} & \textbf{Class-MC} & \textbf{Var-MC} & \textbf{\begin{tabular}[c]{@{}l@{}}$\mu$-MC\\$\mu=0.16$\end{tabular}} & \textbf{\begin{tabular}[c]{@{}l@{}}Noisy\\input\end{tabular}} & \textbf{Class-conv} & \textbf{Class-MC} & \textbf{Var-MC} & \textbf{\begin{tabular}[c]{@{}l@{}}$\mu$-MC\\$\mu=0.16$ \end{tabular}} & \textbf{\begin{tabular}[c]{@{}l@{}}Noisy\\input\end{tabular}} & \textbf{Class-conv} & \textbf{Class-MC} & \textbf{Var-MC} & \textbf{\begin{tabular}[c]{@{}l@{}}$\mu$-MC\\$\mu=0.16$\end{tabular}} & \textbf{\begin{tabular}[c]{@{}l@{}}Noisy\\input\end{tabular}} & \textbf{Class-conv} & \textbf{Class-MC} & \textbf{Var-MC} & \textbf{\begin{tabular}[c]{@{}l@{}}$\mu$-MC\\$\mu=0.16$\end{tabular}} \\ \hline
                  		\multirow{2}{*}{\textbf{-10 dB}} & \textbf{SSE x10\textasciicircum{}4} & 3.68           & 0.0411            & 0.0410           & 0.230           & 0.0499                                                                           & 3.63           & 0.0266            & 0.0281           & 0.0612      & 0.0473                                                                           & 3.55           & 0.0729            & 0.0718           & 0.131      & 0.0894                                                                           & 5.33           & 0.0094            & 0.0097           & 0.367               & 0.0107                                                                           \\ \cline{2-22} 
                  		& \textbf{SSNR}                       & -8..6          & 1.9               & 1.9              & 1.0            & 1.9                                                                              & -8.6           & 2.7               & 2.9              & 2.7      & 2.8                                                                              & -8.5           & 1.5               & 1.5              & 1.3      & 1.5                                                                              & -8.2           & 6.7               & 6.7              &  0.2              & 6.7                                                                              \\ \hline
                  		\multirow{2}{*}{\textbf{-5 dB}}  & \textbf{SSE x10\textasciicircum{}4} & 1.13           & 0.0186            & 0.0187           & 0.124           & 0.0268                                                                           & 1.11           & 0.0128            & 0.0133           & 0.0235      & 0.0180                                                                           & 1.07           & 0.0356            & 0.0360           & 0.0662      & 0.0452                                                                           & 1.68           & 0.0047            & 0.0049           & 0.325               & 0.0083                                                                           \\ \cline{2-22} 
                  		& \textbf{SSNR}                       & -6.8           & 3.5               & 3.5              & 2.5            & 3.5                                                                              & -6.8           & 4.3               & 4.4              & 4.2      & 4.3                                                                              & -6.7           & 2.7               & 2.7              & 2.4      & 2.7                                                                              & -6.3           & 9.1               & 9.1              & 0.8               & 9.1                                                                              \\ \hline
                  		\multirow{2}{*}{\textbf{0 dB}}   & \textbf{SSE x10\textasciicircum{}3} & 3.51           & 0.102             & 0.103            & 0.360      & 0.129                                                                            & 3.35           & 0.076             & 0.082            & 0.133               & 0.10                                                                             & 3.21           & 0.191             & 0.197            & 0.298      & 0.236                                                                            & 5.28           & 0.036             & 0.037            & 2.47               & 0.070                                                                            \\ \cline{2-22} 
                  		& \textbf{SSNR}                       & -4.2           & 5.3               & 5.3              & 4.3      & 5.3                                                                              & -4.3           & 5.9               & 5.9              & 5.6      & 5.9                                                                              & -4.1           & 4.2               & 4.2              & 3.8      & 4.2                                                                              & -3.6           & 10.8              & 10.8             &  2.1              & 10.8                                                                             \\ \hline
                  		\multirow{2}{*}{\textbf{5 dB}}   & \textbf{SSE x10\textasciicircum{}3} & 1.08           & 0.067             & 0.069            & 0.134      & 0.075                                                                            & 0.999          & 0.055             & 0.062            & 0.083               & 0.069                                                                            & 0.956          & 0.115             & 0.121            & 0.153               & 0.128                                                                            & 1.66           & 0.033             & 0.034            &  1.45              & 0.076                                                                            \\ \cline{2-22} 
                  		& \textbf{SSNR}                       & -1.1           & 7.3               & 7.3              & 6.3      & 7.3                                                                              & -1.1           & 7.4               & 7.4              & 7.0               & 7.4                                                                              & -1.0           & 5.8               & 5.8              & 5.3               & 5.8                                                                              & -0.3           & 12.1              & 12.1             &  4.6              & 12.0                                                                             \\ \hline
                  		\multirow{2}{*}{\textbf{10 dB}}  & \textbf{SSE x10\textasciicircum{}2} & 3.30           & 0.52              & 0.54             & 0.78      & 0.55                                                                             & 2.95           & 0.48              & 0.50             & 0.64               & 0.52                                                                             & 2.84           & 0.87              & 0.87             & 0.90               & 0.83                                                                             & 5.18           & 0.33              & 0.34             &  5.01              & 0.51                                                                             \\ \cline{2-22} 
                  		& \textbf{SSNR}                       & 2.5            & 9.1               & 9.1              & 8.1      & 9.1                                                                              & 2.5            & 8.9               & 8.9              & 8.5               & 8.9                                                                              & 2.6            & 7.5               & 7.5              & 7.0               & 7.5                                                                              & 3.3            & 12.9              & 12.9             & 7.5               & 12.8                                                                             \\ \hline
                  	\end{tabular}
                  }
                  	\label{table3}
                  \end{table*}    
   
Figure \ref{fig4} shows the correlation between the predictive variance and the squared error (SE) of the estimated output frames for all the five MC models, for speech corrupted with white noise. The uncertainty is computed by taking the trace of the covariance matrix of each frame \cite{kendall2016modelling}. The plots show the weakening of the correlation between  the SE and model uncertainty as the SNR improves. The correlation is strong for -10 and -5 dB and is weak for the values of SNR 0, 5 and 10 dB. This variation could be due to the fact that  the DNN is less adapted to lower SNRs and highly adapted to high SNRs. This needs further exploration. This matches with our results, since we find that there is not much improvement over the class-conv and class-MC as the SNR increases. However, the values are still comparable to the same. This observation matches with the observations in \cite{kendall2016modelling}, that the test data which are far from training set are likely to be more uncertain as the network is less adapted to them.

\subsubsection{Observations}
Tables \ref{table2} and \ref{table3} shows that Var-MC gives really poor performance for seen noises like factory2, m109, leopard, babble and volvo. $\mu$-MC algorithm compensates for this performance drop by using per frame predictive variance threshold $\mu$ to select between Var-MC and class-MC.

The threshold is selected based on the experiments on a validation set of 30 files from TIMIT corrupted with seen noises factory 2, m109, leopard, babble and volvo noises  and an unseen pink noise at SNRs -10, -5 , 0, 5 and 10 dB. For our experiments this threshold is set to be $\mu=0.16$. 


\subsection{$\mu$-MC}

Tables \ref{table2} and \ref{table3} shows the performance improvements of $\mu$-MC algorithm over class-conv and class-MC in terms of SSE and SSNR for  unseen noises pink, white and factory 1 and for seen noises factory 2, m109, leopard, babble and volvo. It can be observed that $\mu$-MC gives superior performance in most of the unseen noise cases, especially at lower SNRs. The algorithm also compensates for the poor performance of Var-MC algorithm for seen noises.

\begin{figure}[t!]%
  	
  	
  	\centering

  	\includegraphics[height=7cm,width=9.7cm]{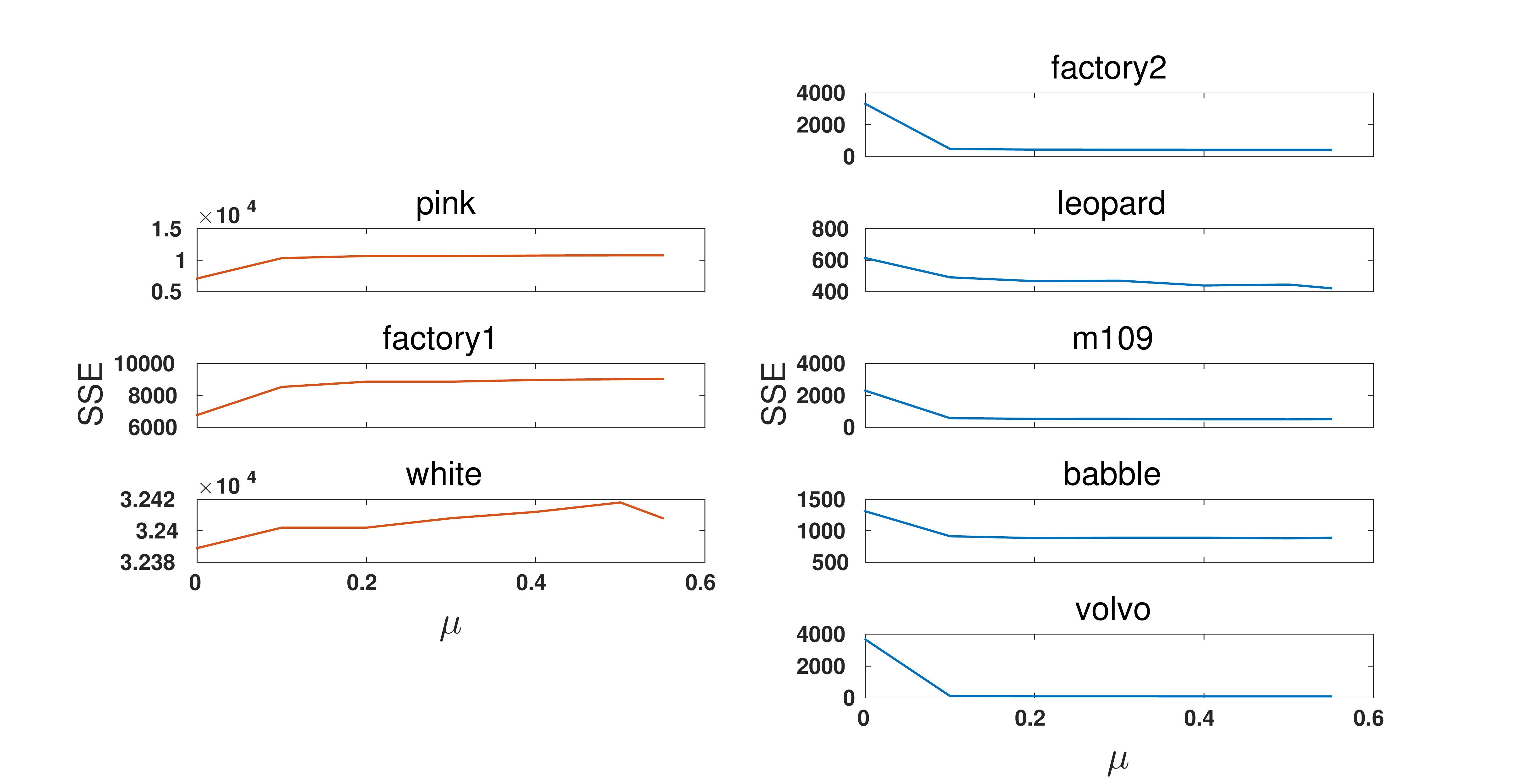}

  
  	\caption{Variation of SSE with predictive variance $\mu$ for -10dB}
  		\label{fig5}	
  \end{figure}

\begin{figure*}
	\resizebox{1.05\linewidth}{7.5cm}{
	\begin{tabular}{p{.07\textwidth}p{.19\textwidth}p{.19\textwidth}p{.19\textwidth}p{.19\textwidth}p{.19\textwidth}}
		& \hspace{1em}Factory2 model & \hspace{-.7em}Leopard model & \hspace{-1.5em}
		M109 model & \hspace{-2em}Babble model & \hspace{-3em}Volvo model\\
		-10dB  
		
		&\hspace{-1em}\IncG[width=.2\textwidth,height=.13\textwidth]{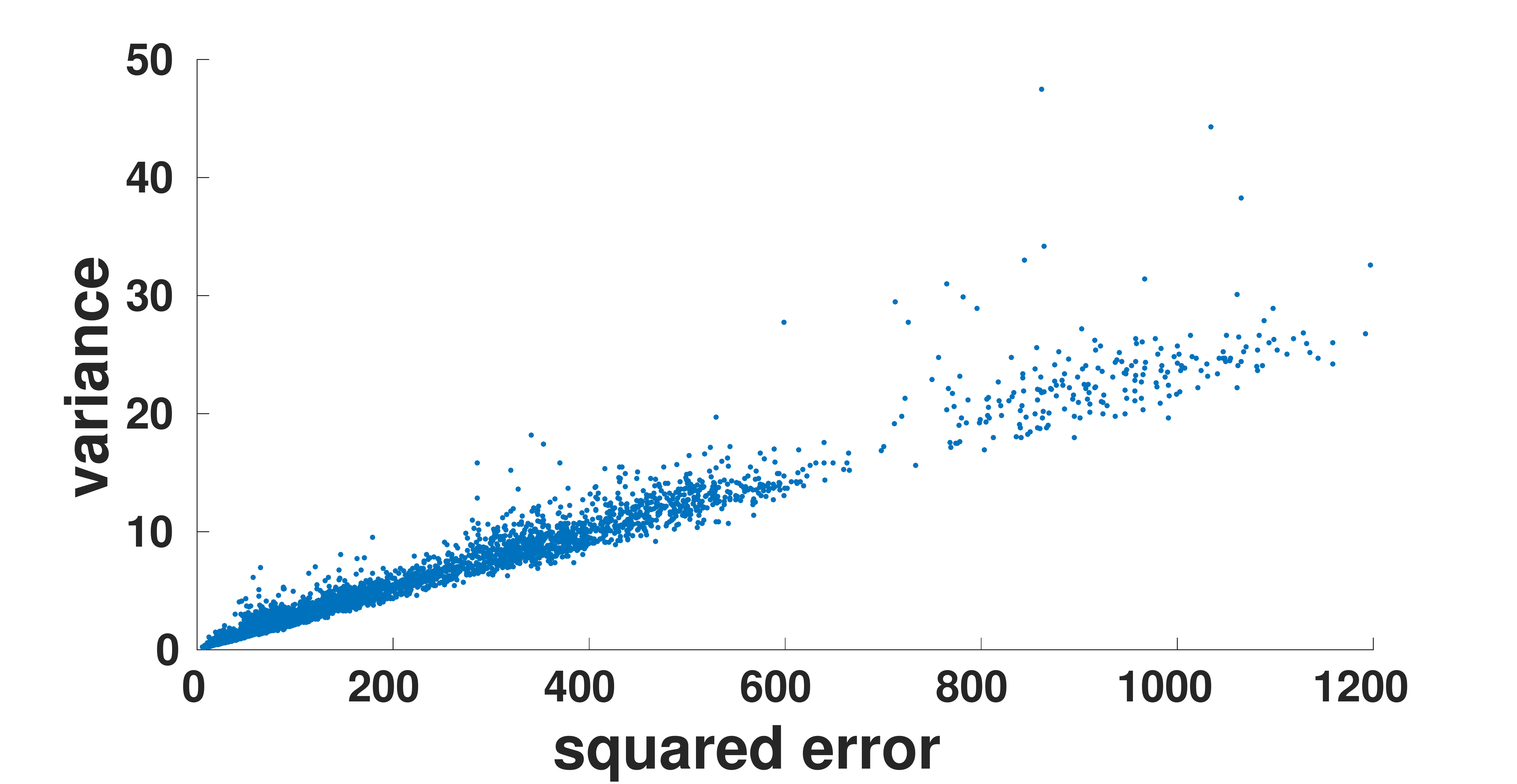}
		&\hspace{-3em}\IncG[width=.2\textwidth,height=.13\textwidth]{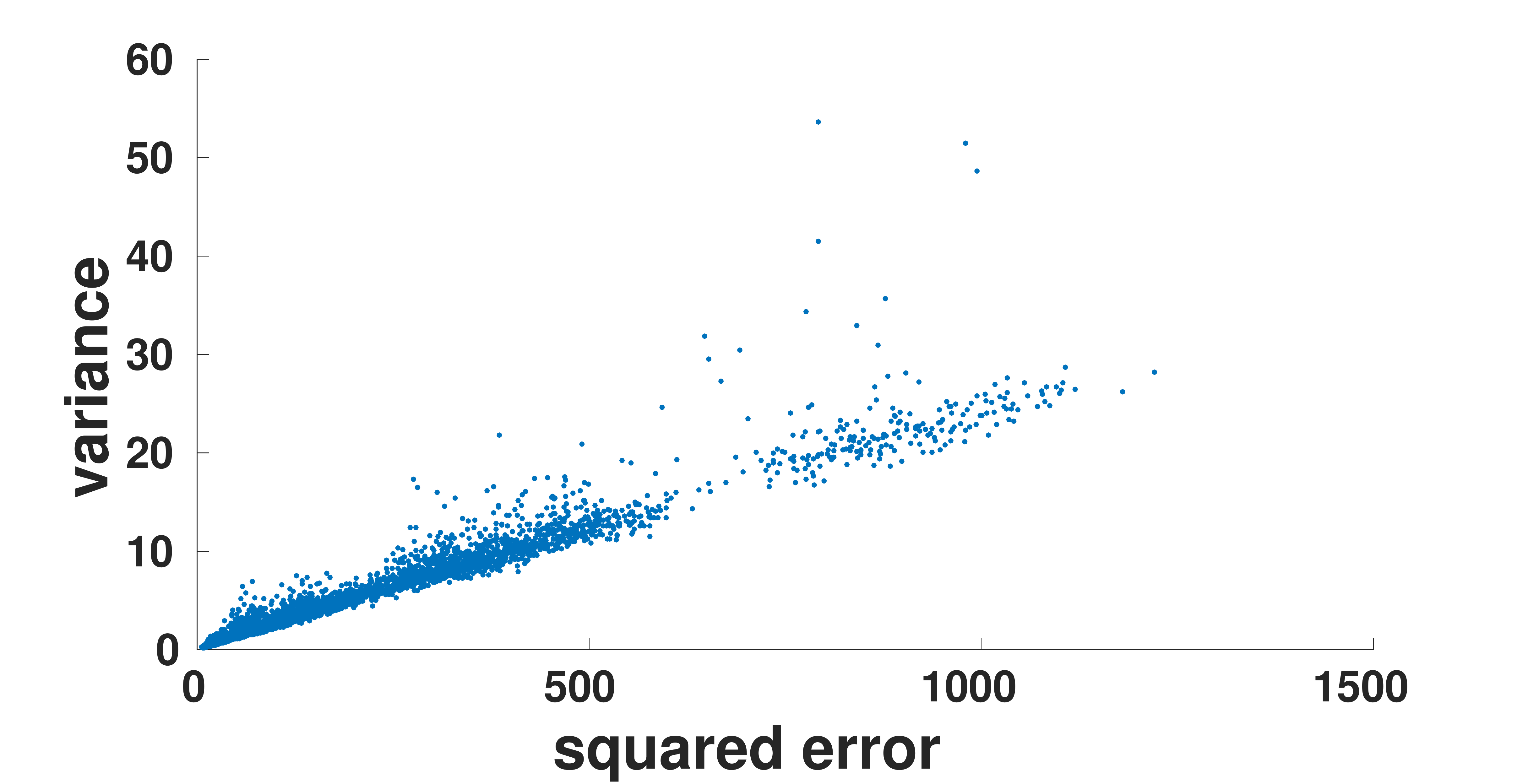}
		&\hspace{-3em}\IncG[width=.2\textwidth,height=.13\textwidth]{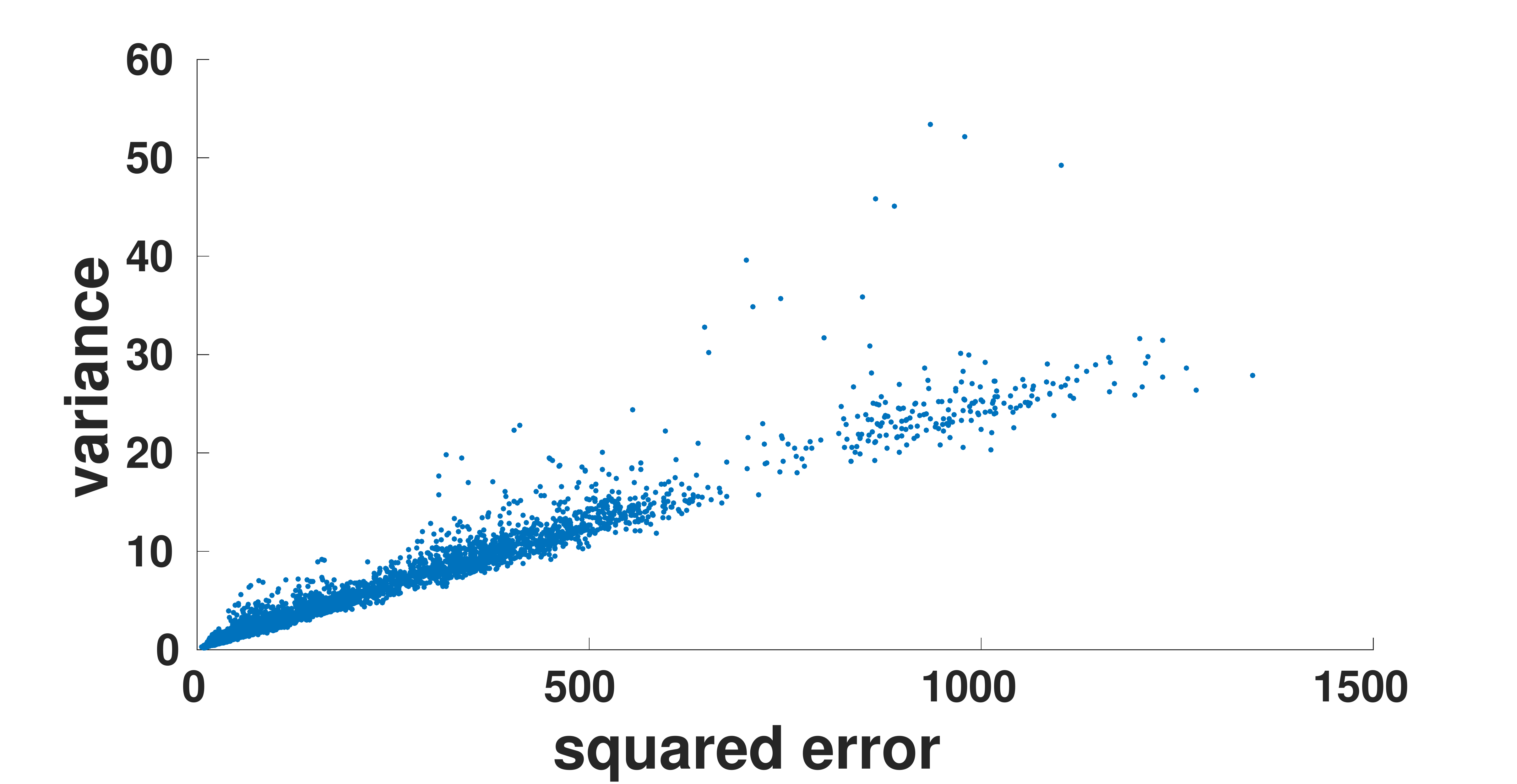}
		&\hspace{-3em}\IncG[width=.2\textwidth,height=.13\textwidth]{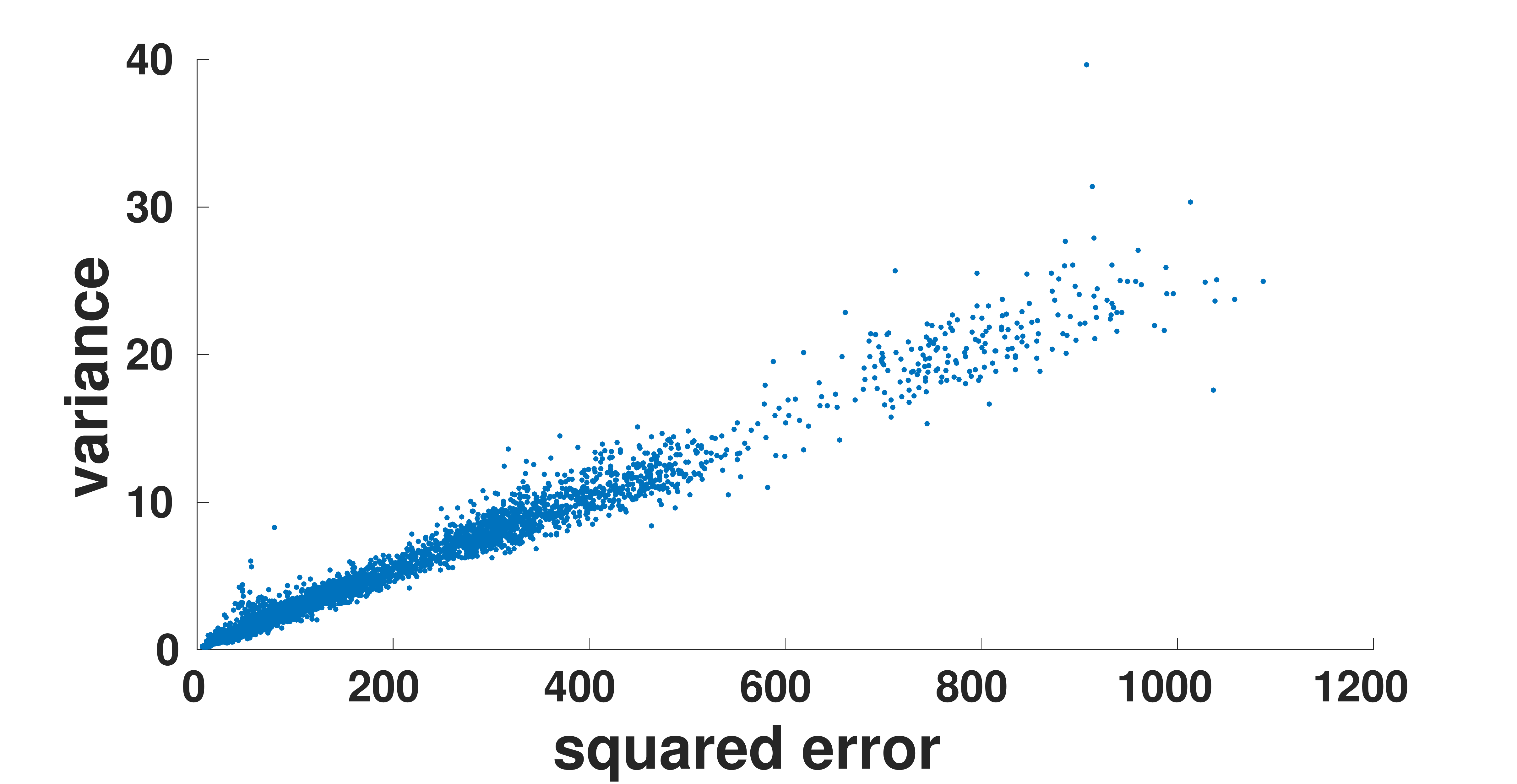}
		&\hspace{-4em}\IncG[width=.2\textwidth,height=.13\textwidth]{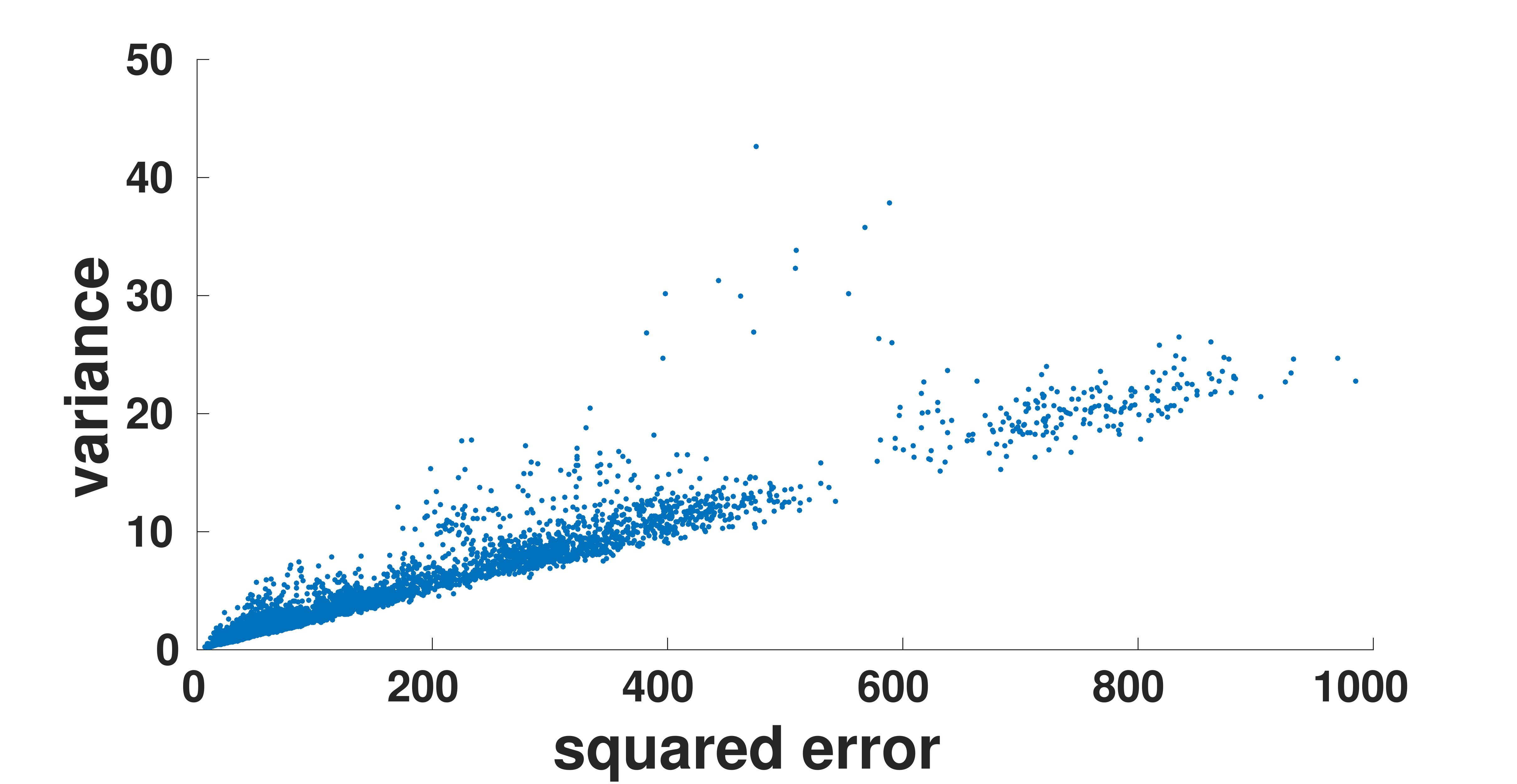}\\
		-5dB 
		&\hspace{-1em}\IncG[width=.2\textwidth,height=.13\textwidth]{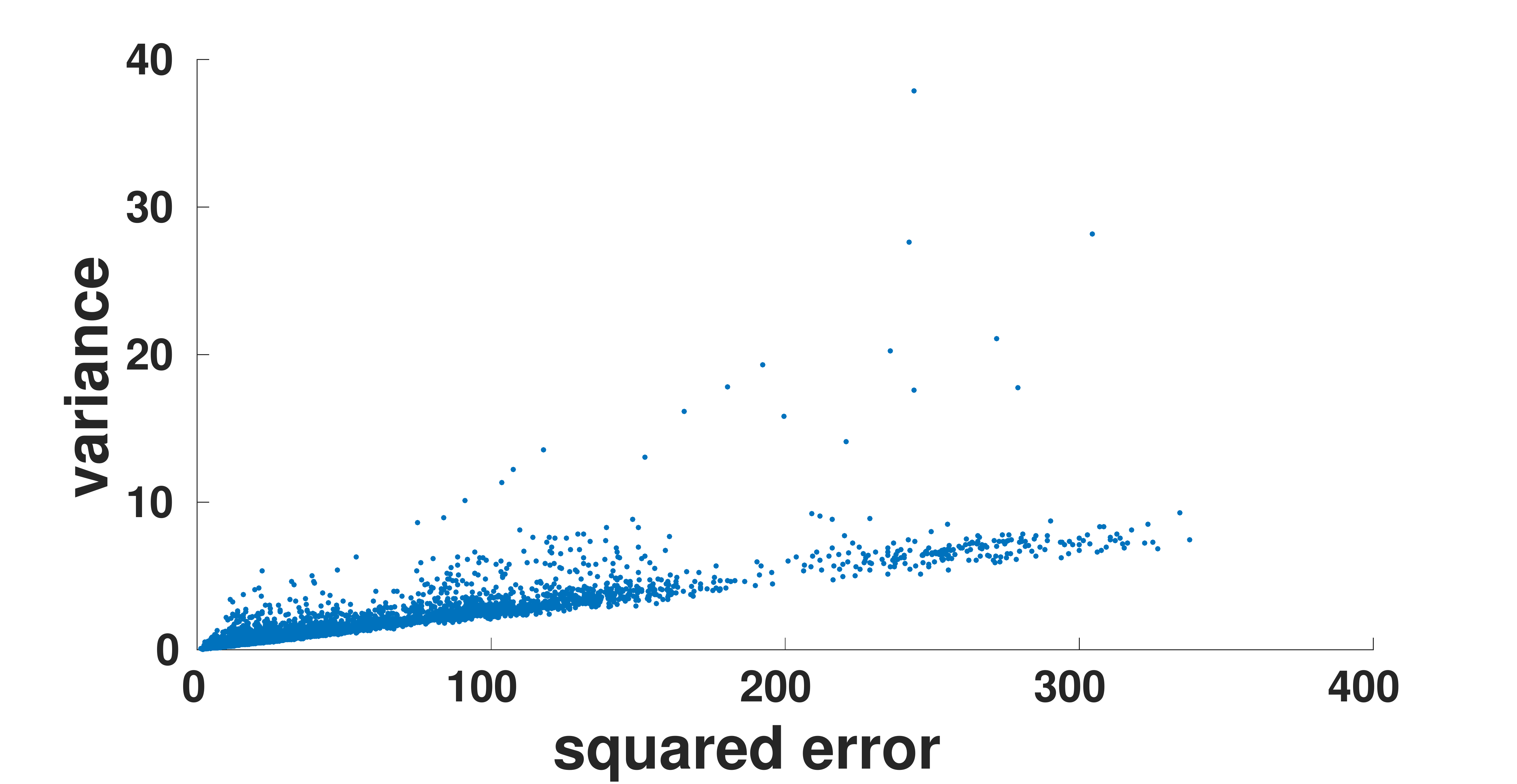}
		&\hspace{-3em}\IncG[width=.2\textwidth,height=.12\textwidth]{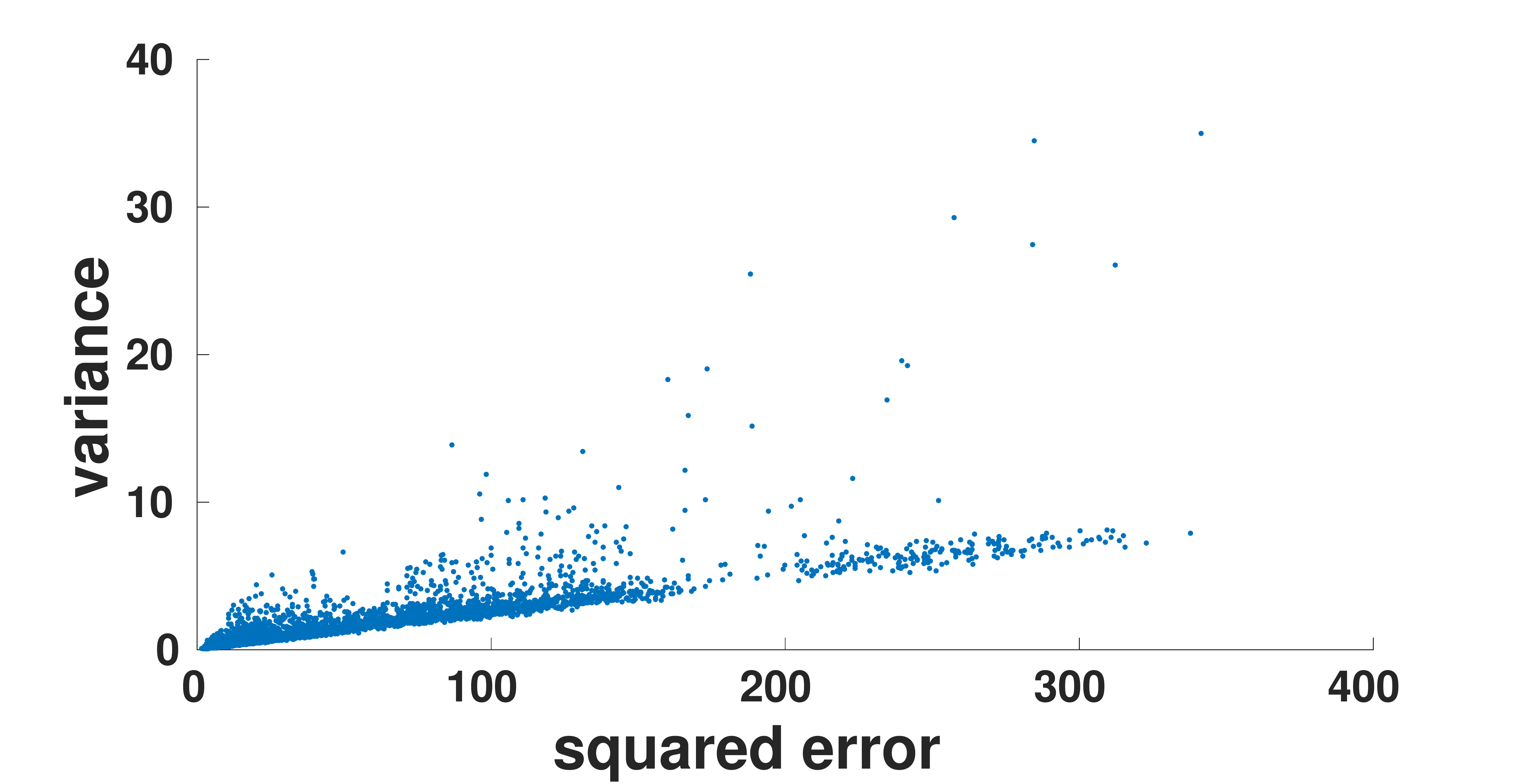}
		&\hspace{-3em}\IncG[width=.2\textwidth,height=.12\textwidth]{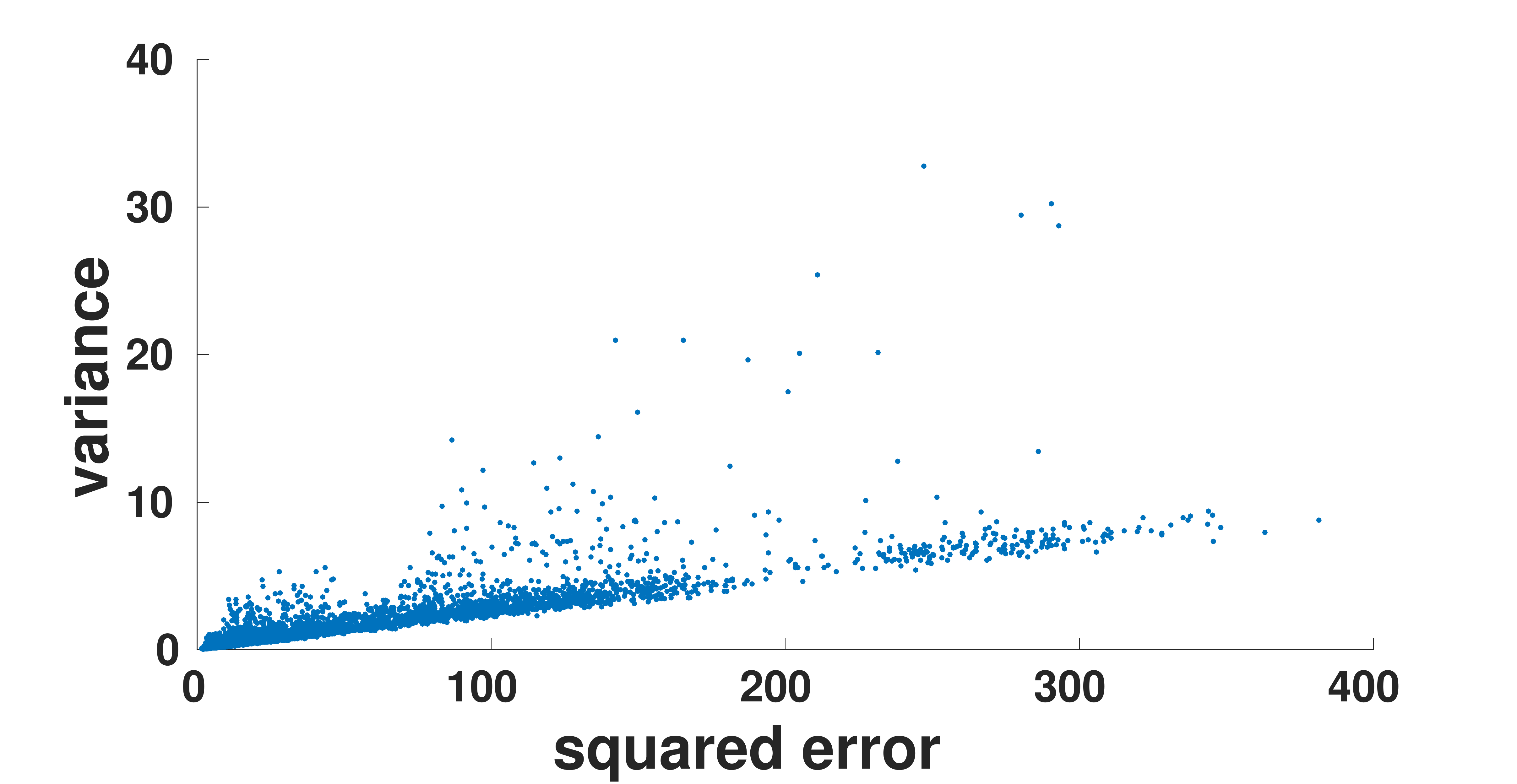}
		&\hspace{-3em}\IncG[width=.2\textwidth,height=.12\textwidth]{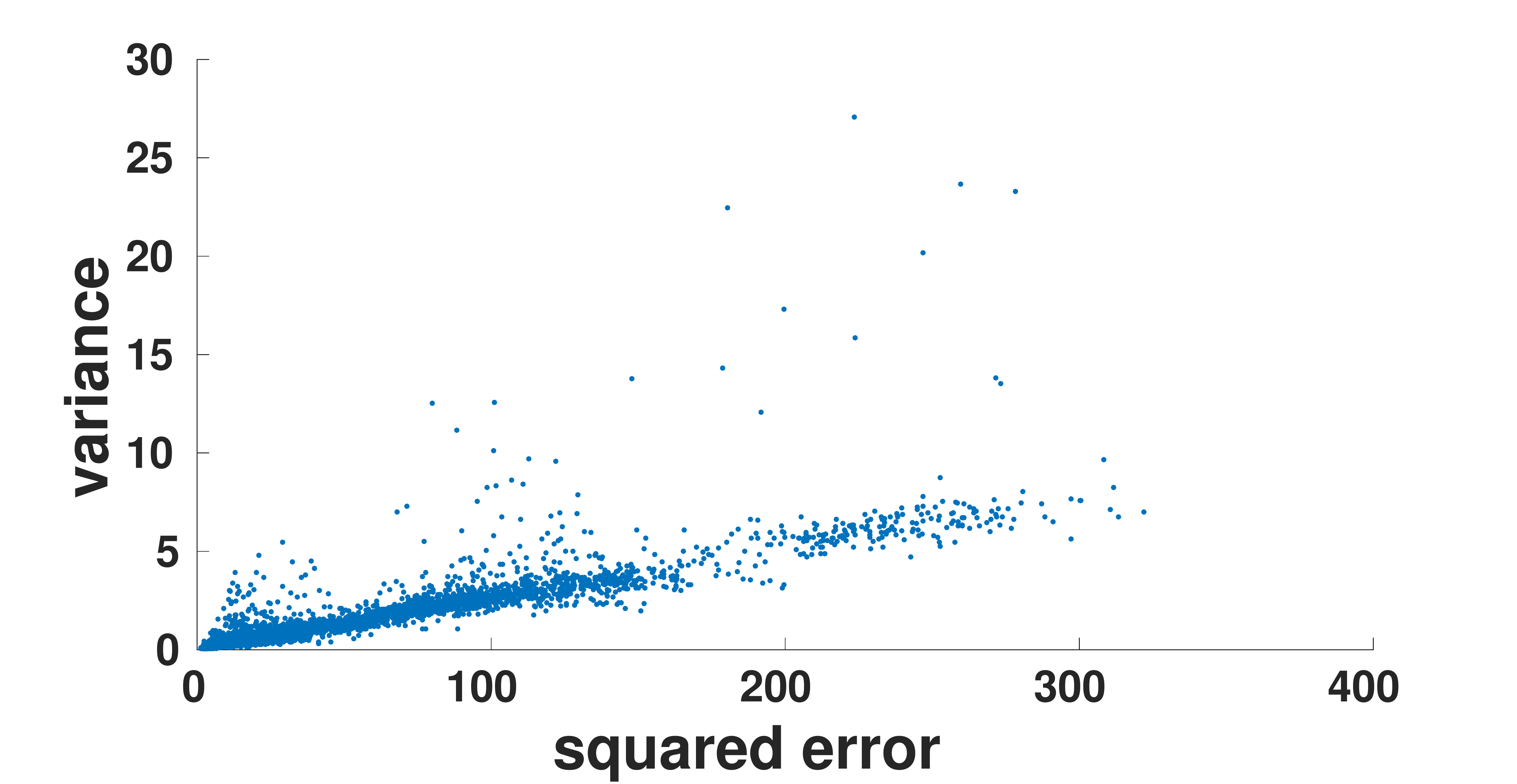}
		&\hspace{-4em}\IncG[width=.2\textwidth,height=.12\textwidth]{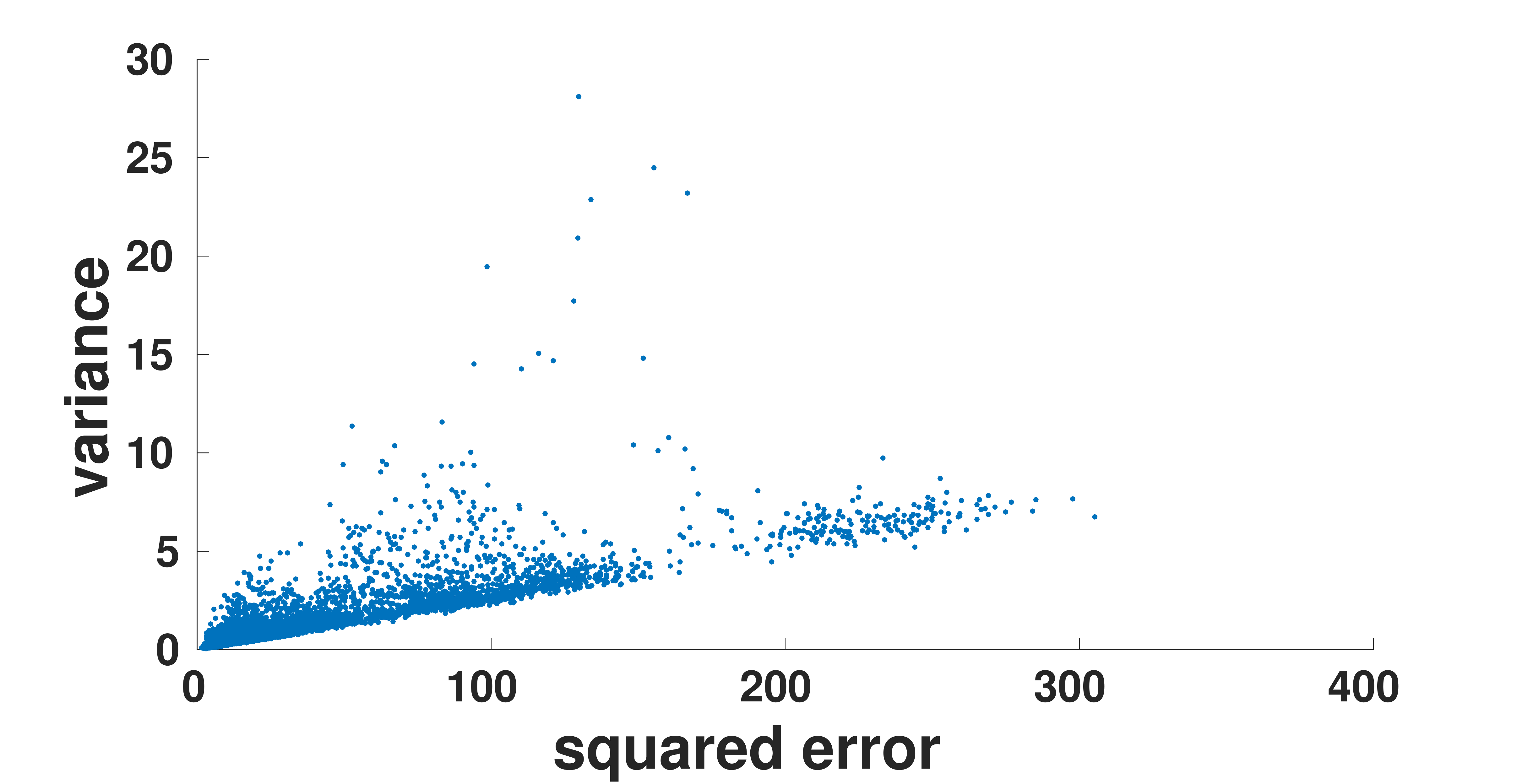}\\
		0dB 
		&\hspace{-1em}\IncG[width=.2\textwidth,height=.12\textwidth]{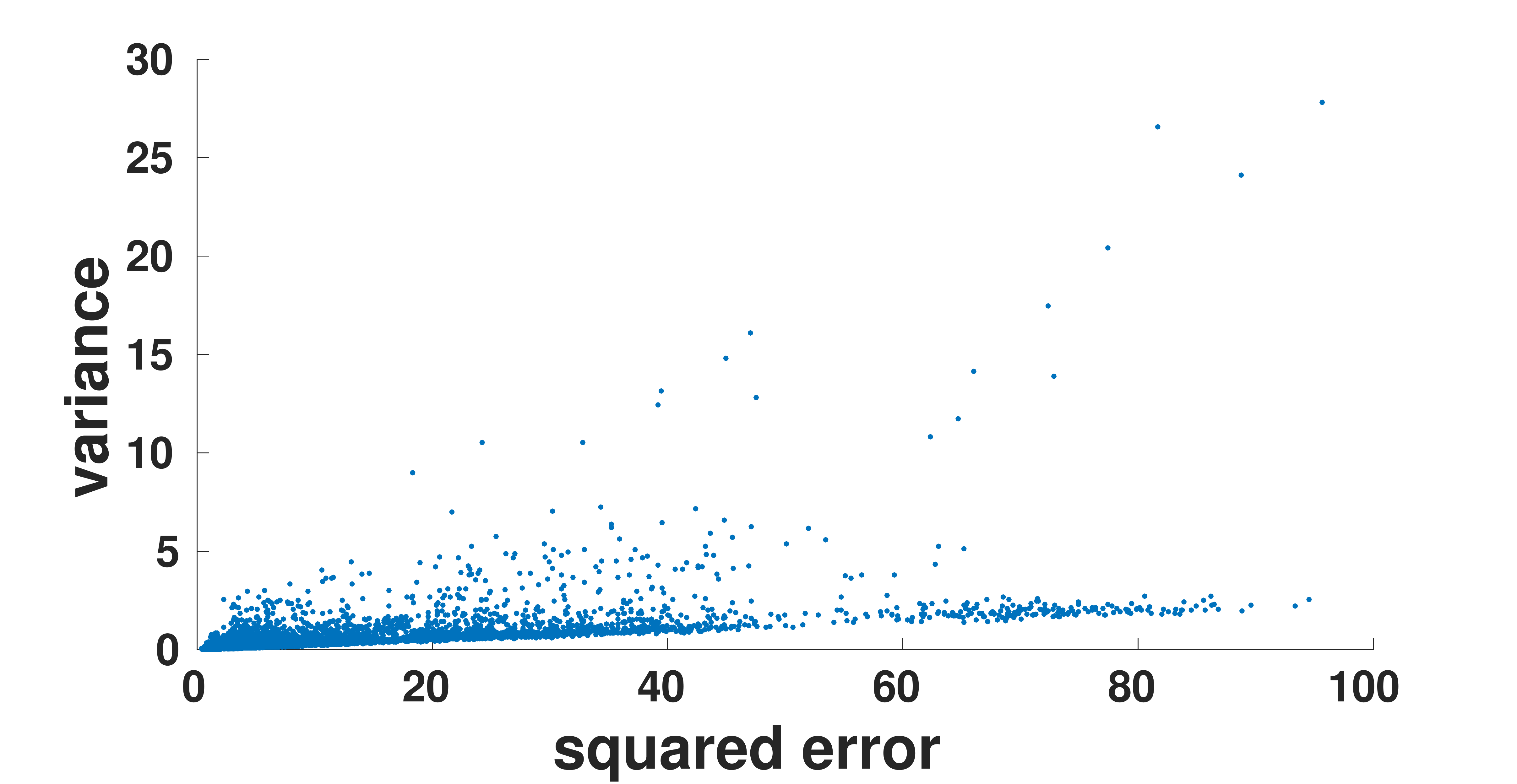}
		&\hspace{-3em}\IncG[width=.2\textwidth,height=.12\textwidth]{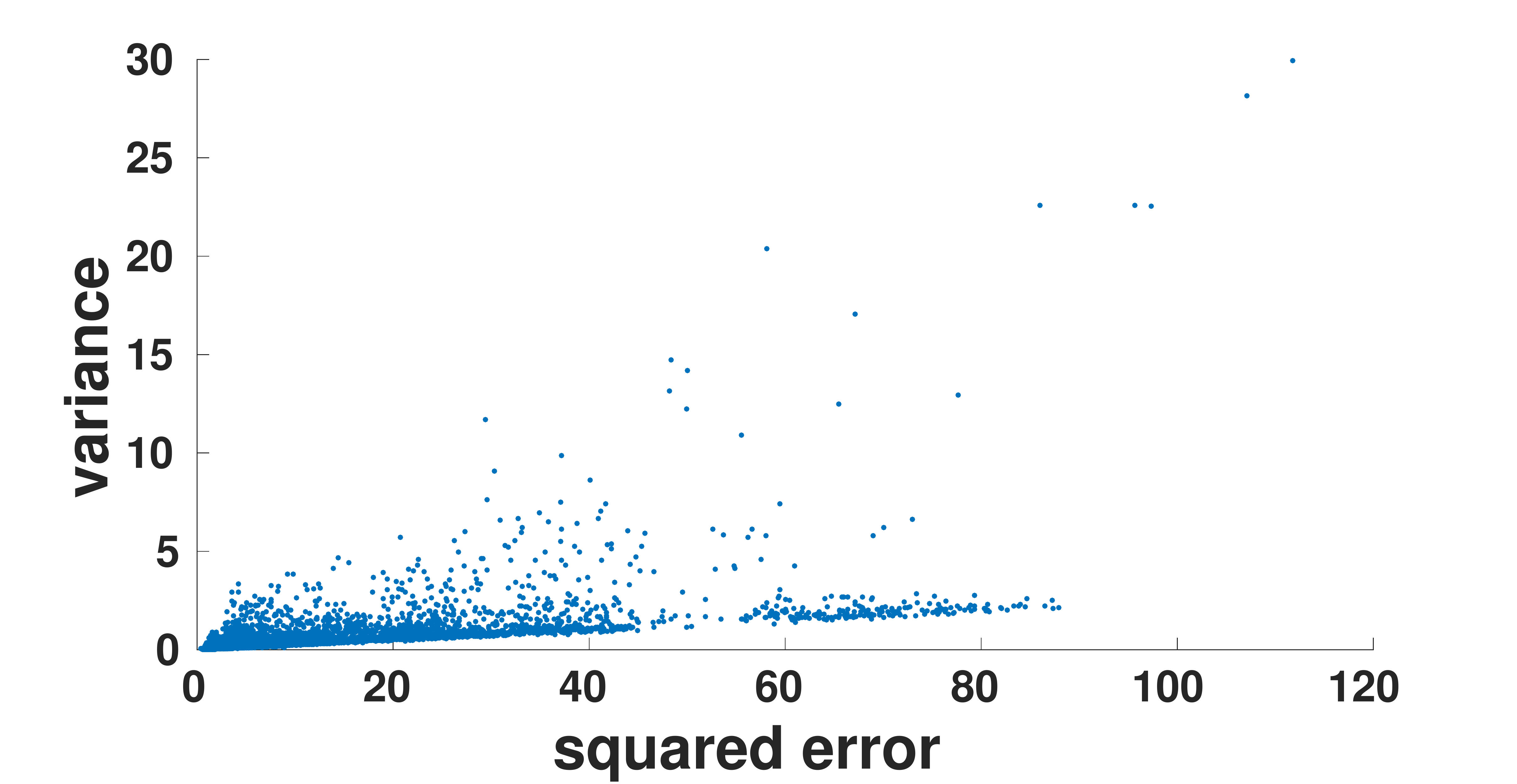}
		&\hspace{-3em}\IncG[width=.2\textwidth,height=.12\textwidth]{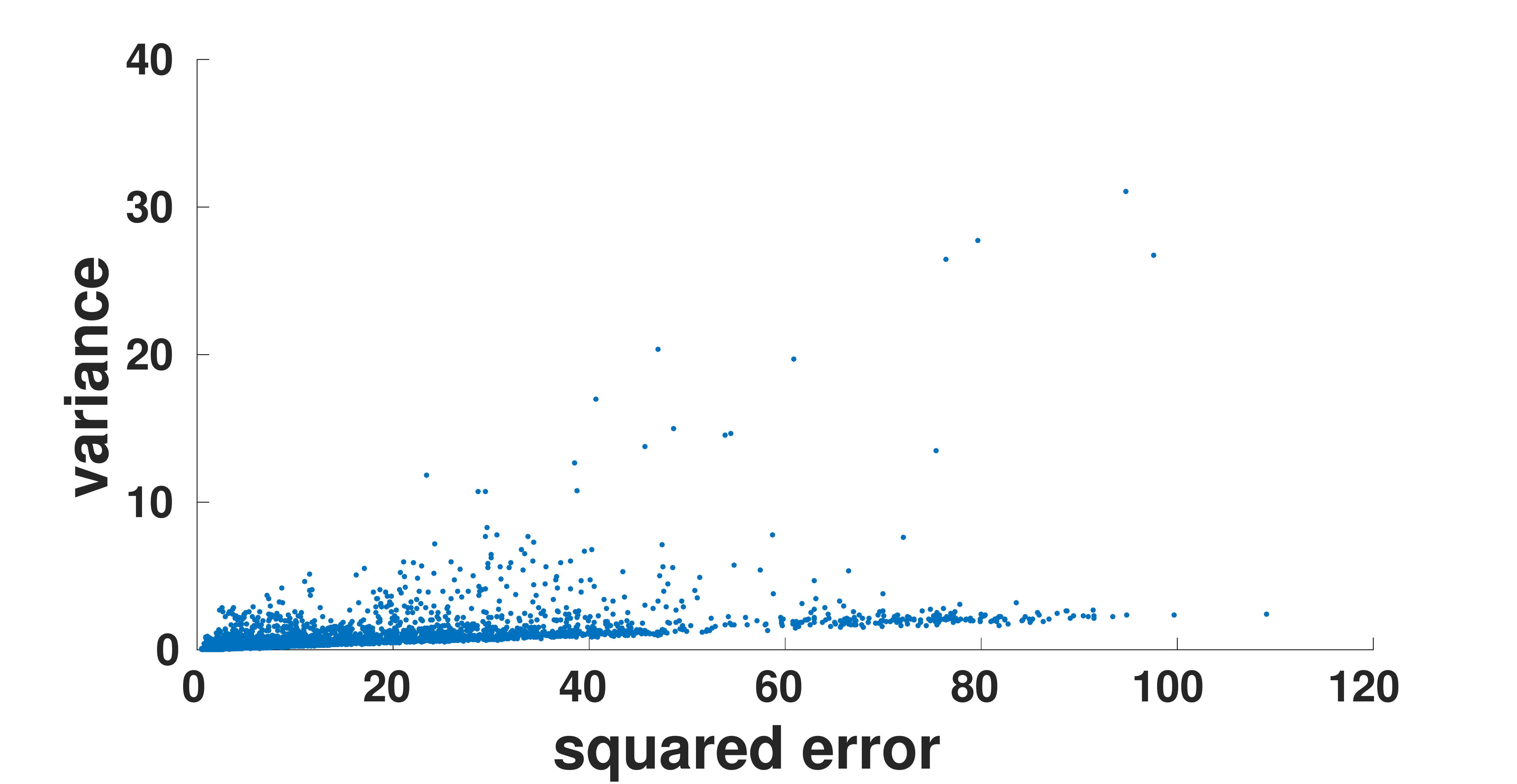}
		&\hspace{-3em}\IncG[width=.2\textwidth,height=.12\textwidth]{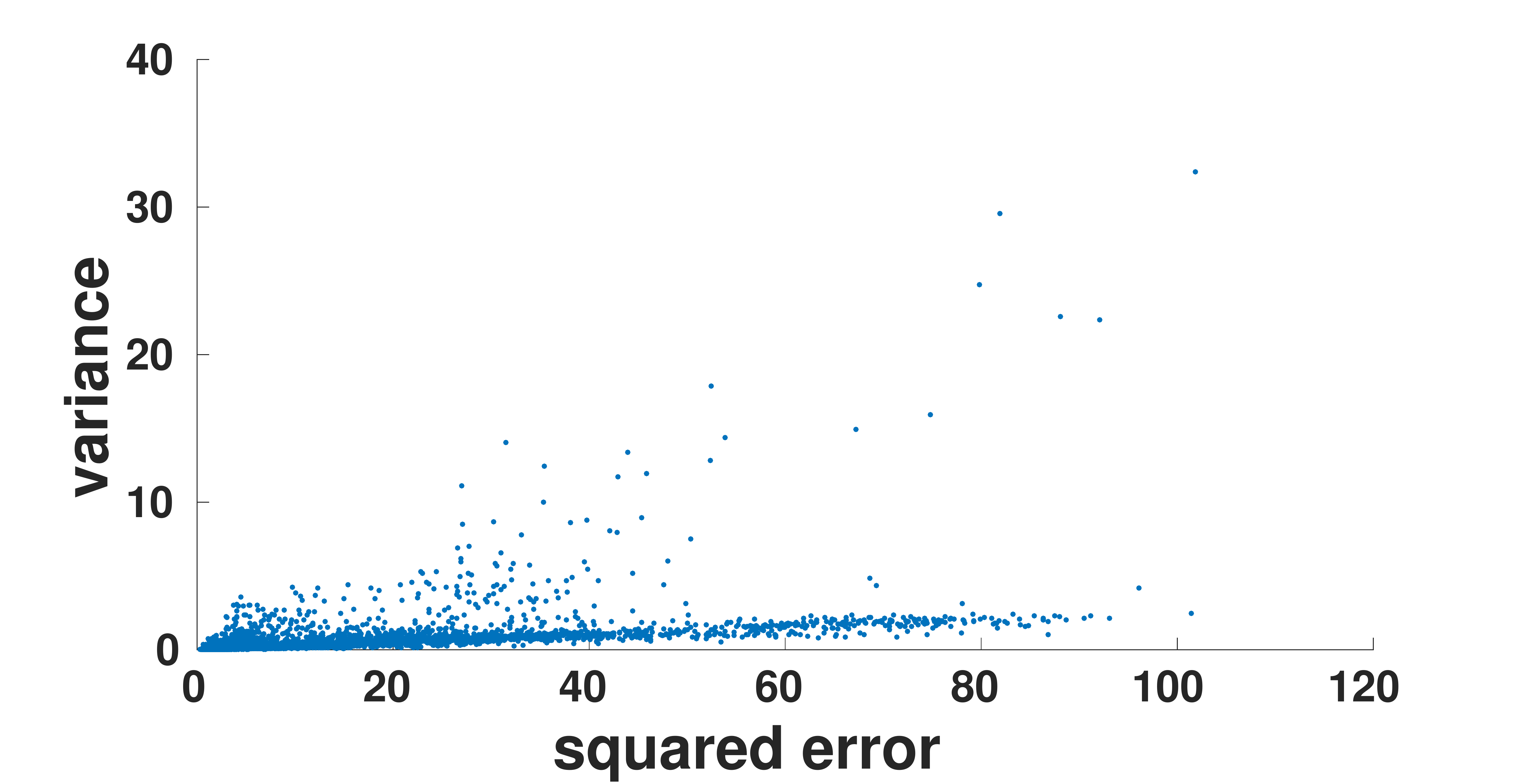}
		&\hspace{-4em}\IncG[width=.2\textwidth,height=.12\textwidth]{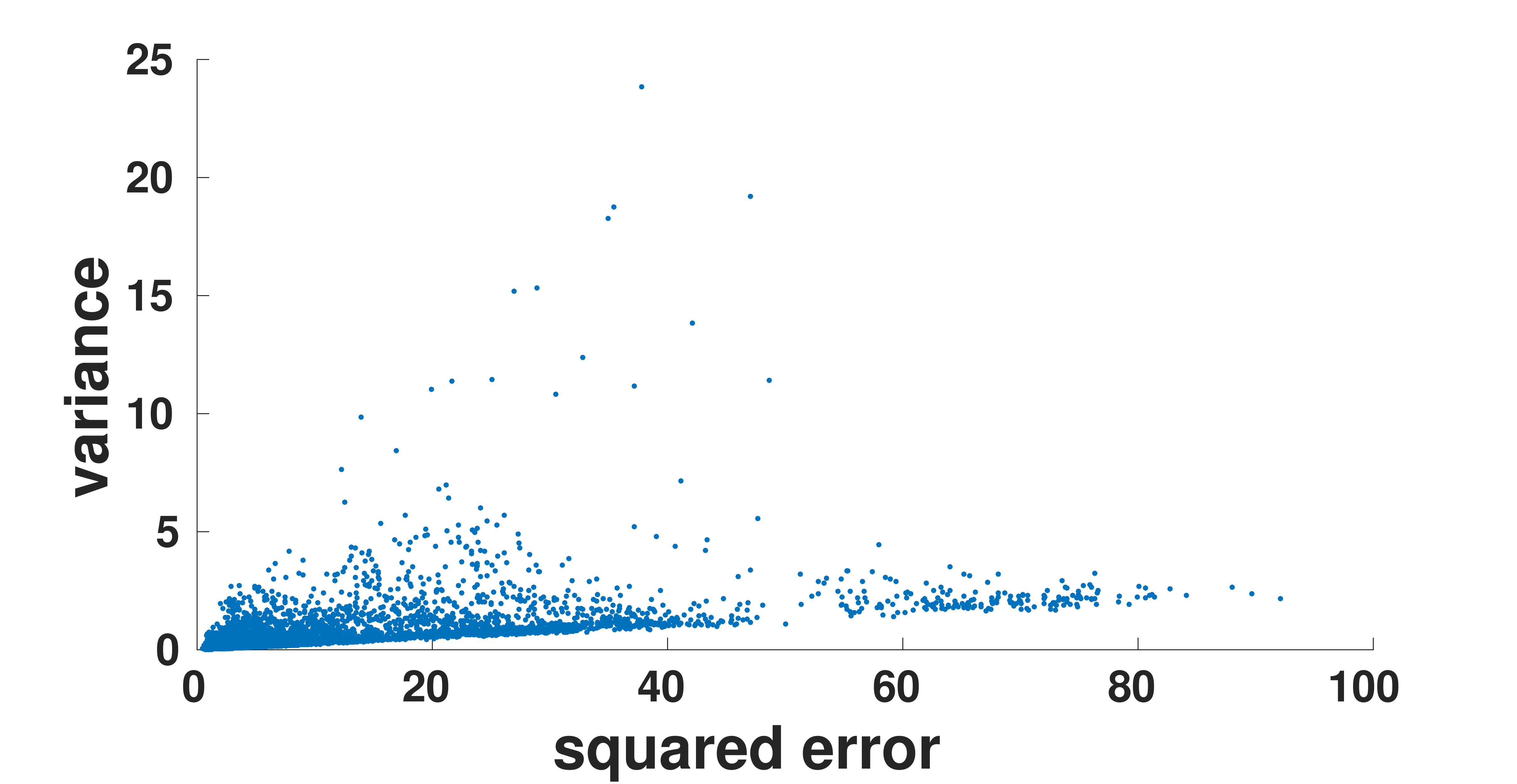}\\
		5dB 
		&\hspace{-1em}\IncG[width=.2\textwidth,height=.12\textwidth]{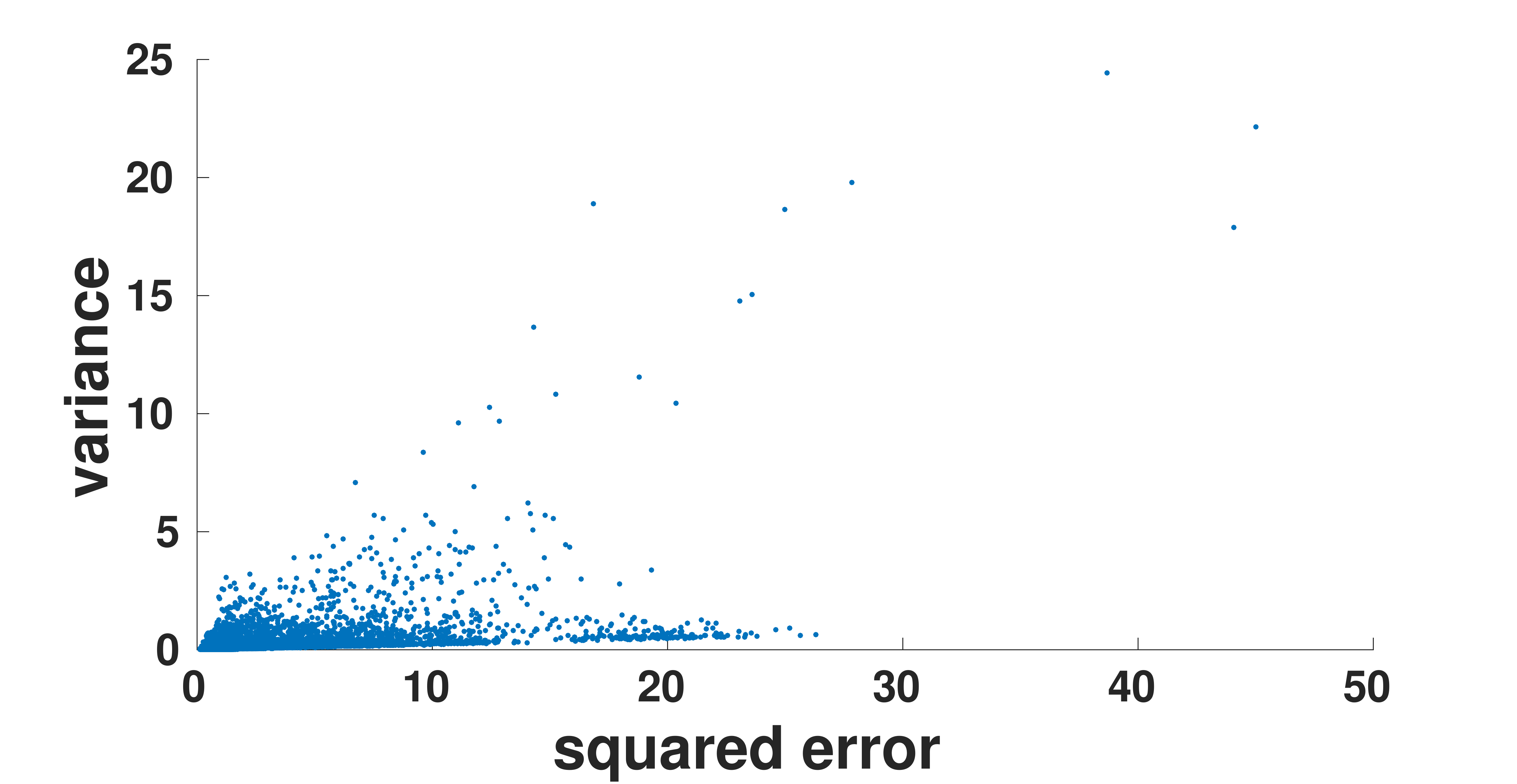}
		&\hspace{-3em}\IncG[width=.2\textwidth,height=.12\textwidth]{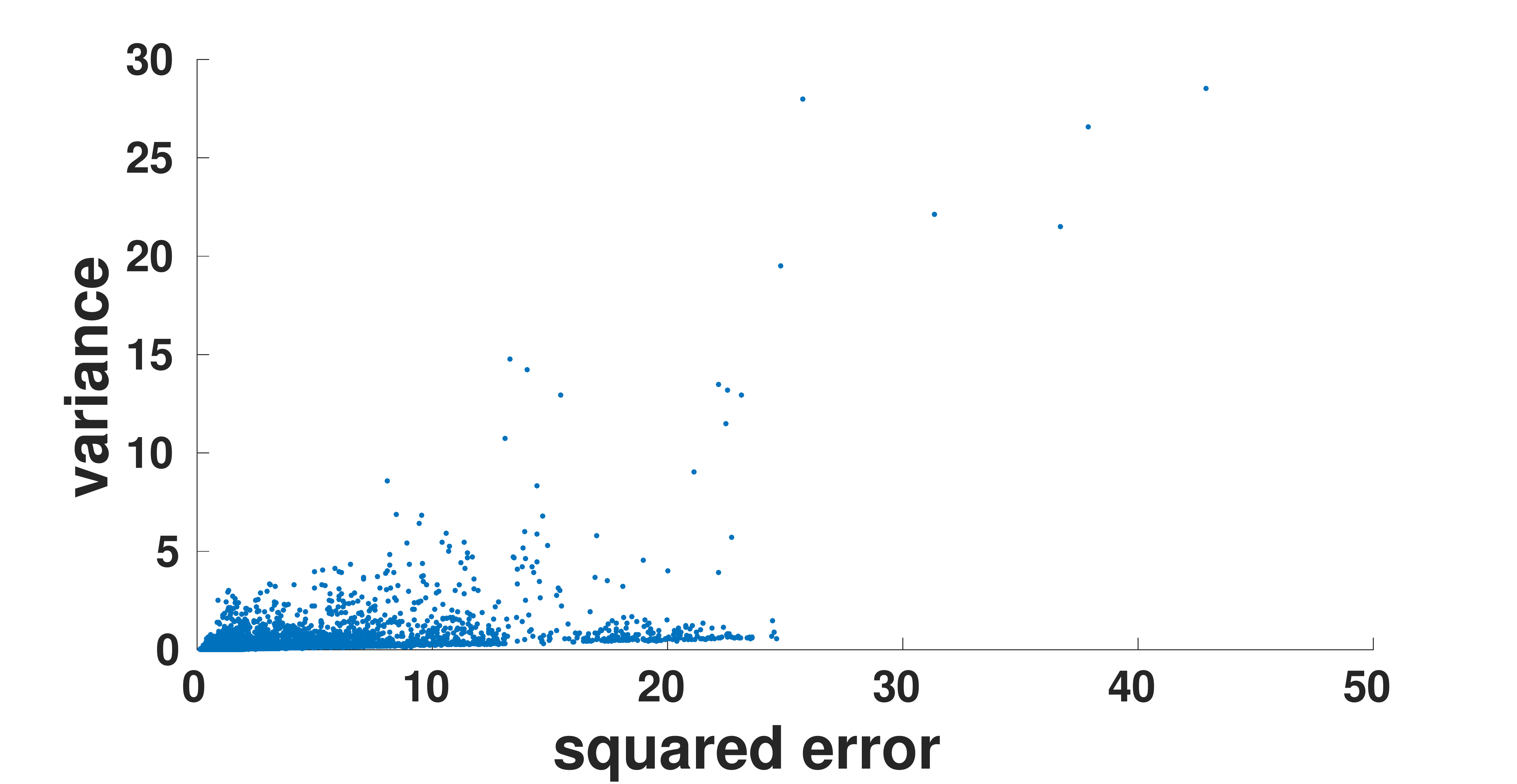}
		&\hspace{-3em}\IncG[width=.2\textwidth,height=.12\textwidth]{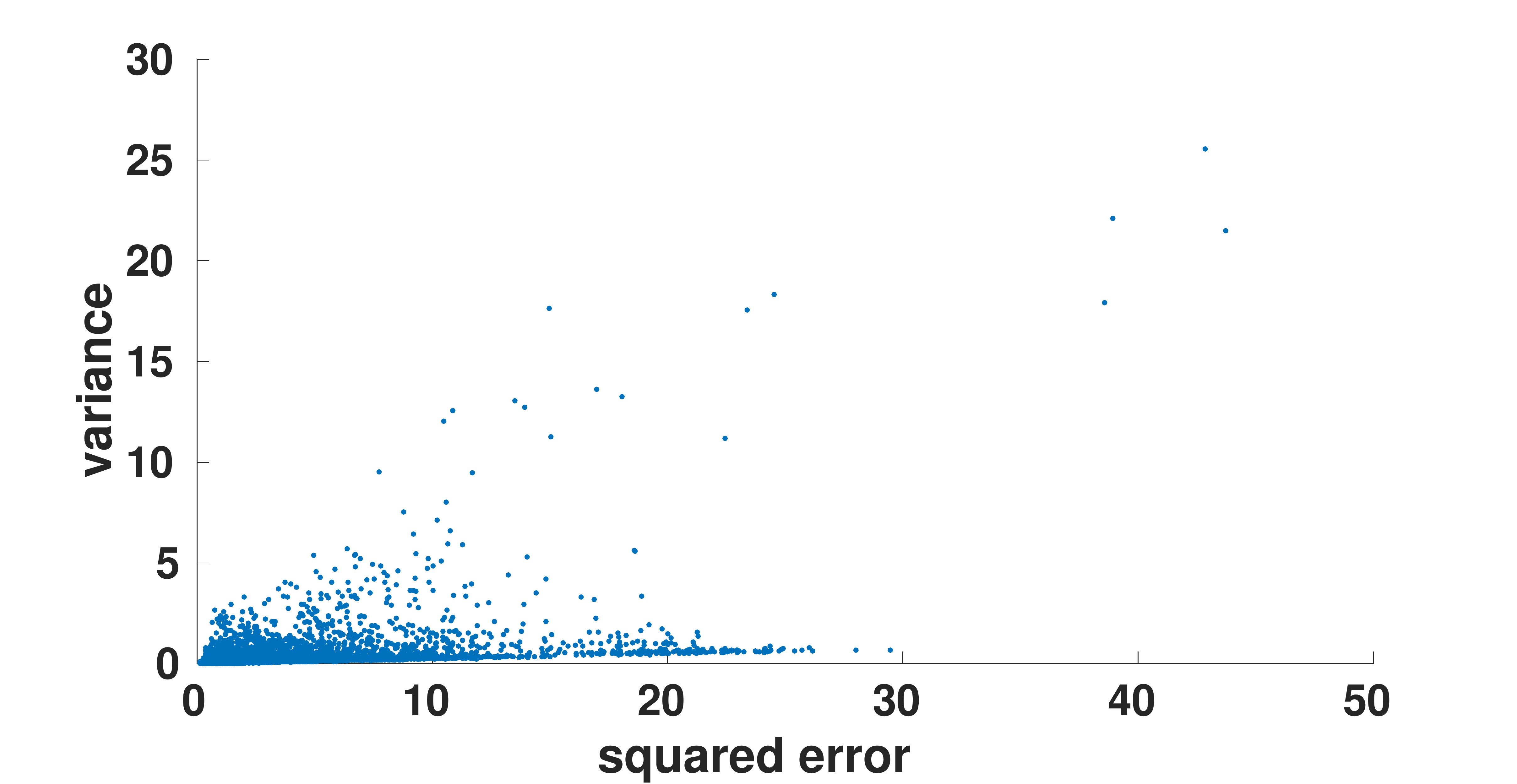}
		&\hspace{-3em}\IncG[width=.2\textwidth,height=.12\textwidth]{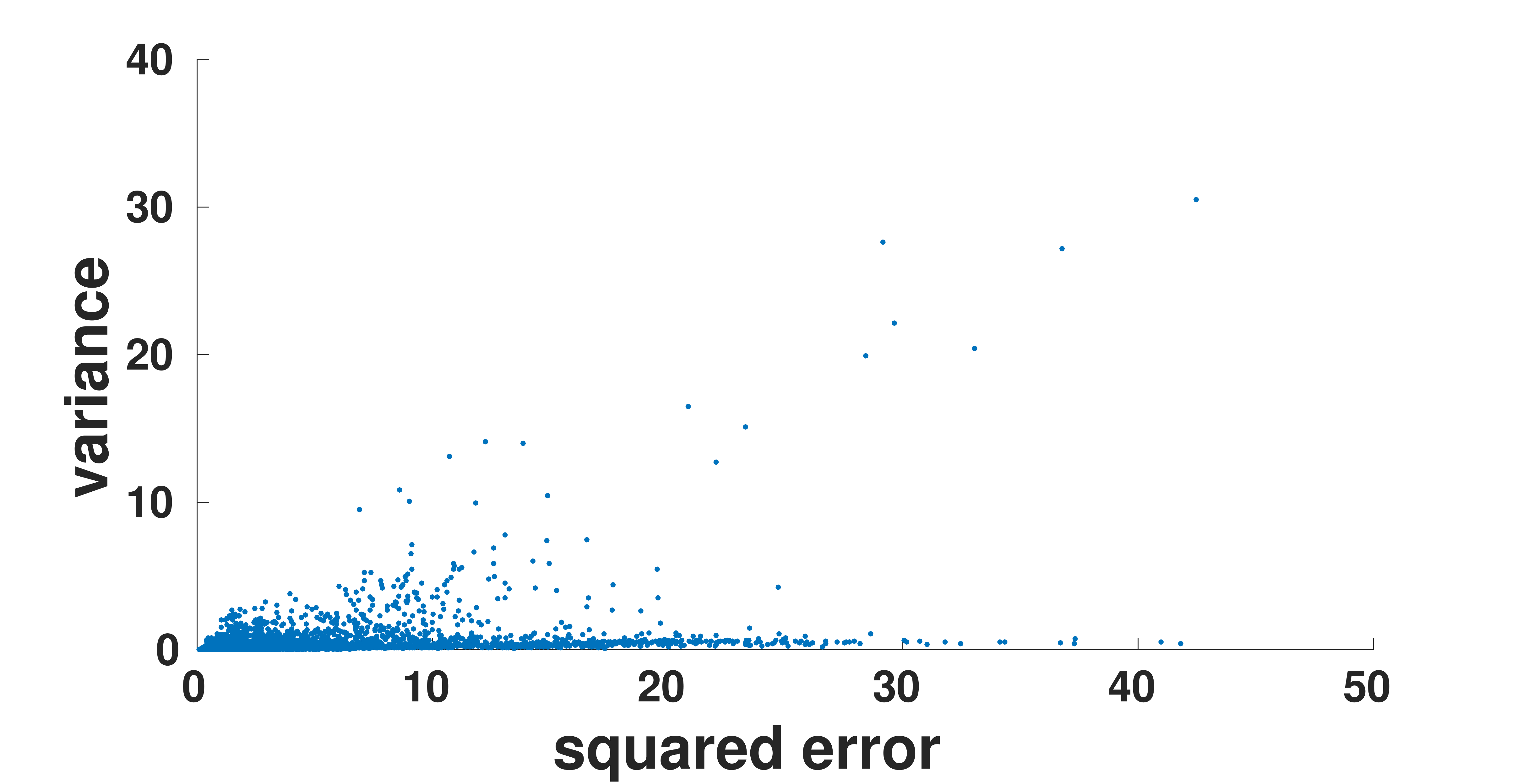}
		&\hspace{-4em}\IncG[width=.2\textwidth,height=.12\textwidth]{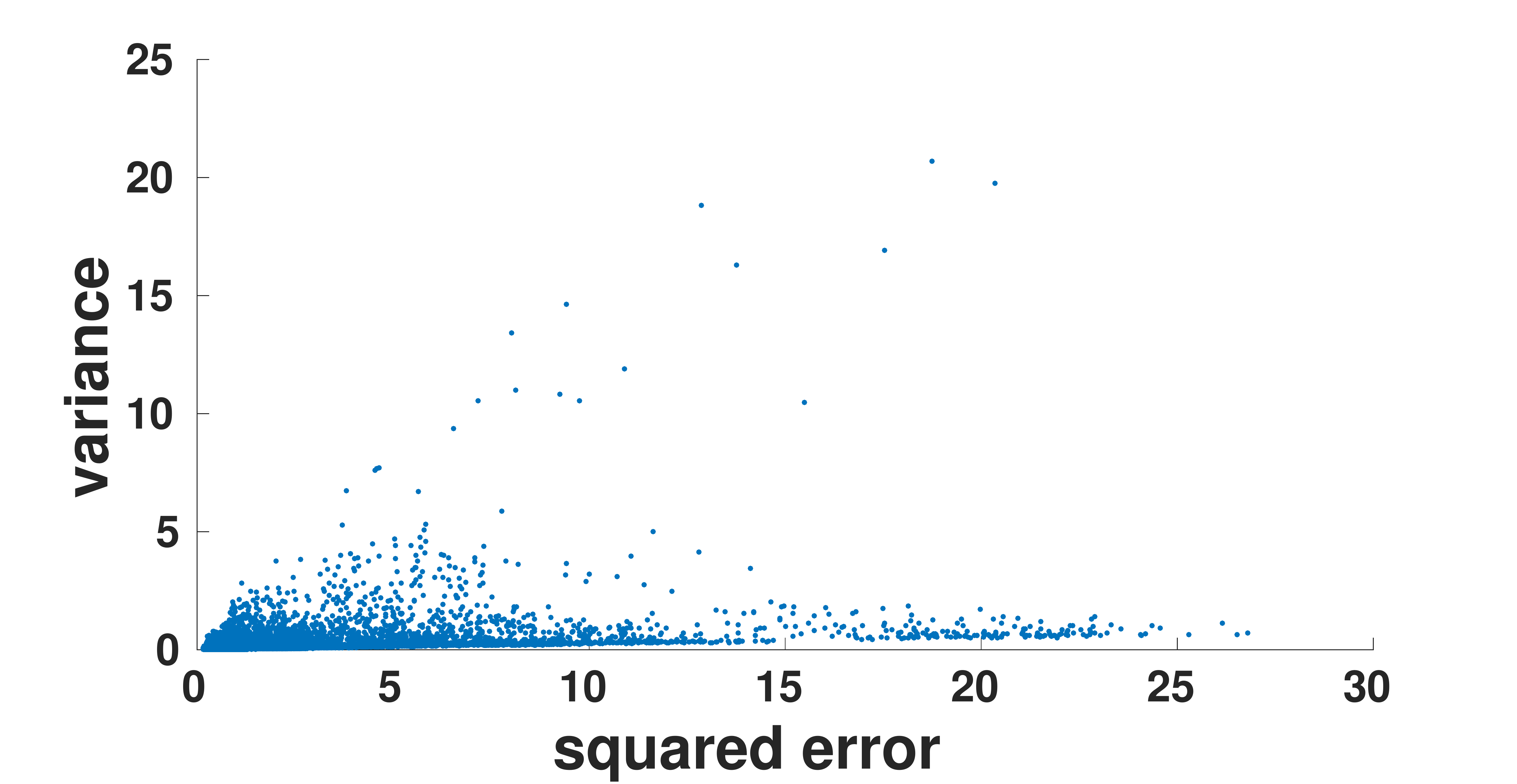}\\
		10dB 
		&\hspace{-1em}\IncG[width=.2\textwidth,height=.12\textwidth]{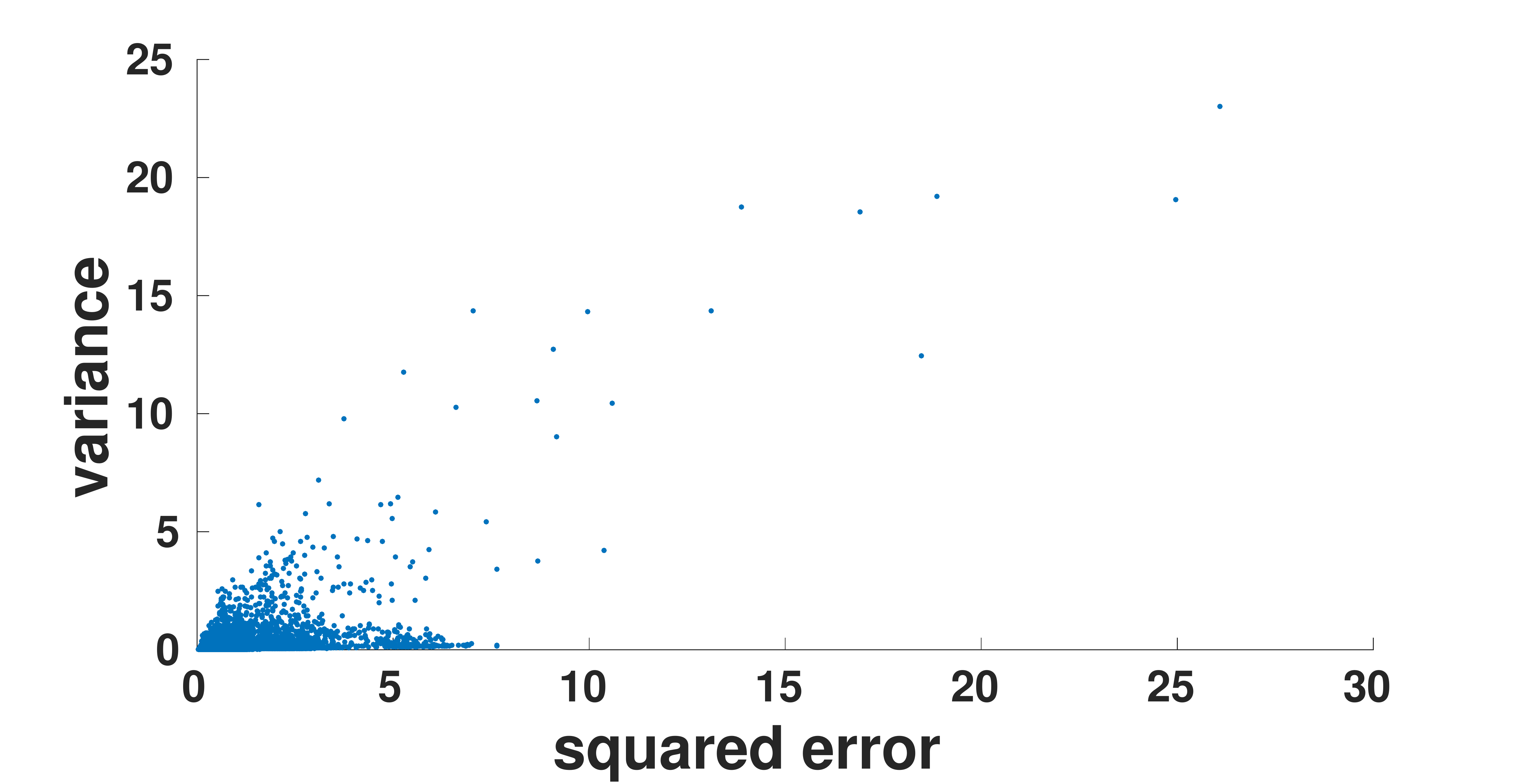}
		&\hspace{-3em}\IncG[width=.2\textwidth,height=.12\textwidth]{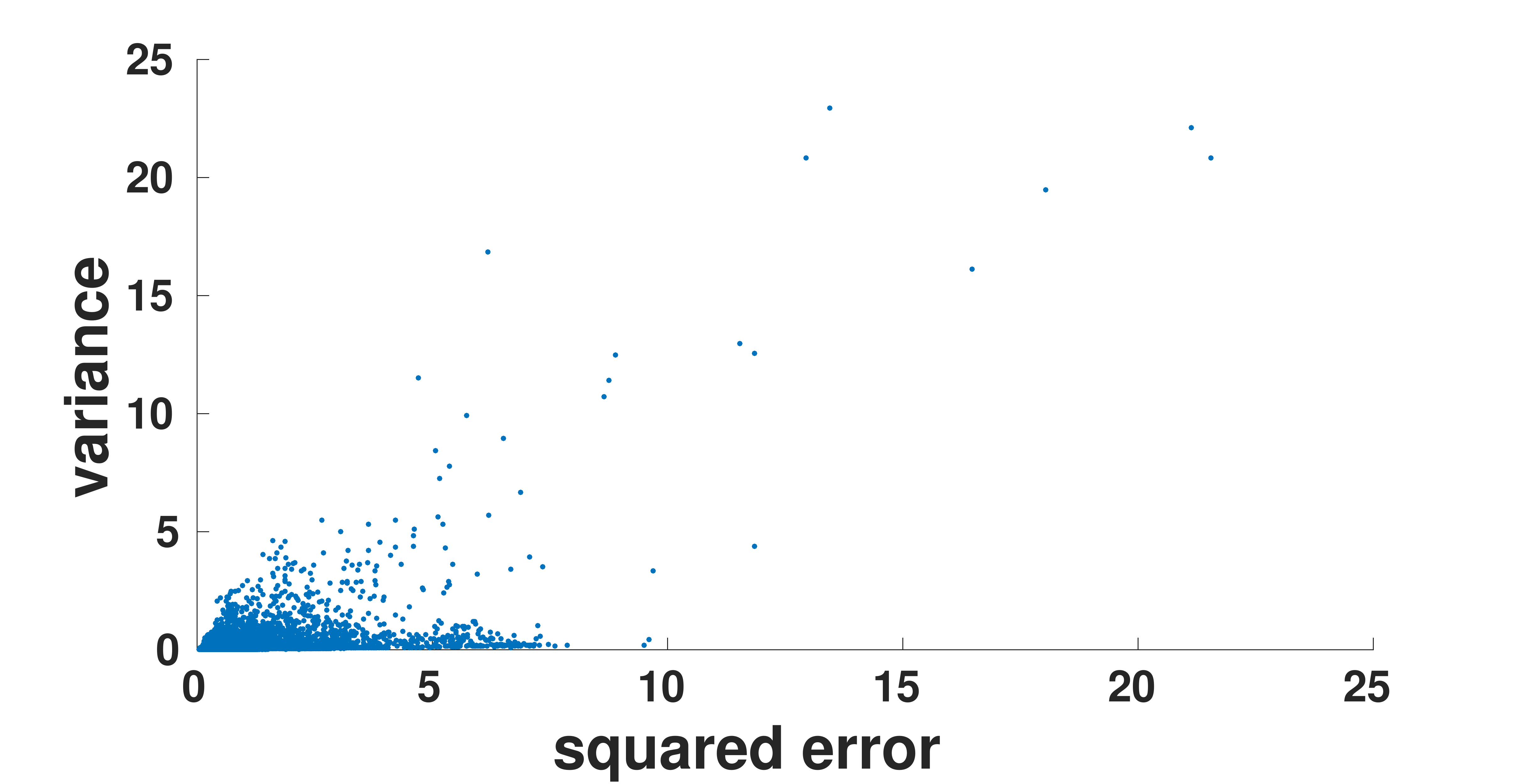}
		&\hspace{-3em}\IncG[width=.2\textwidth,height=.12\textwidth]{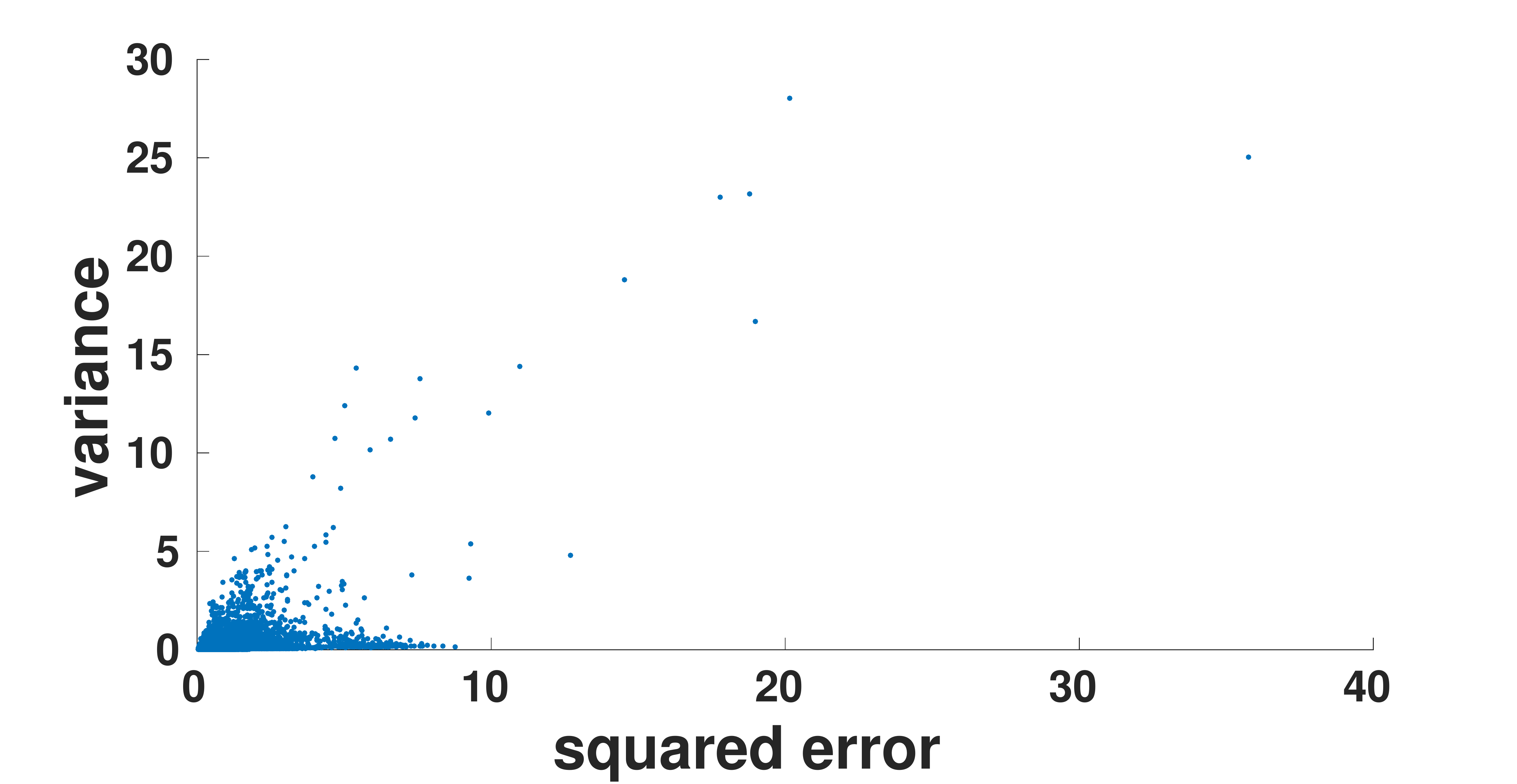}
		&\hspace{-3em}\IncG[width=.2\textwidth,height=.12\textwidth]{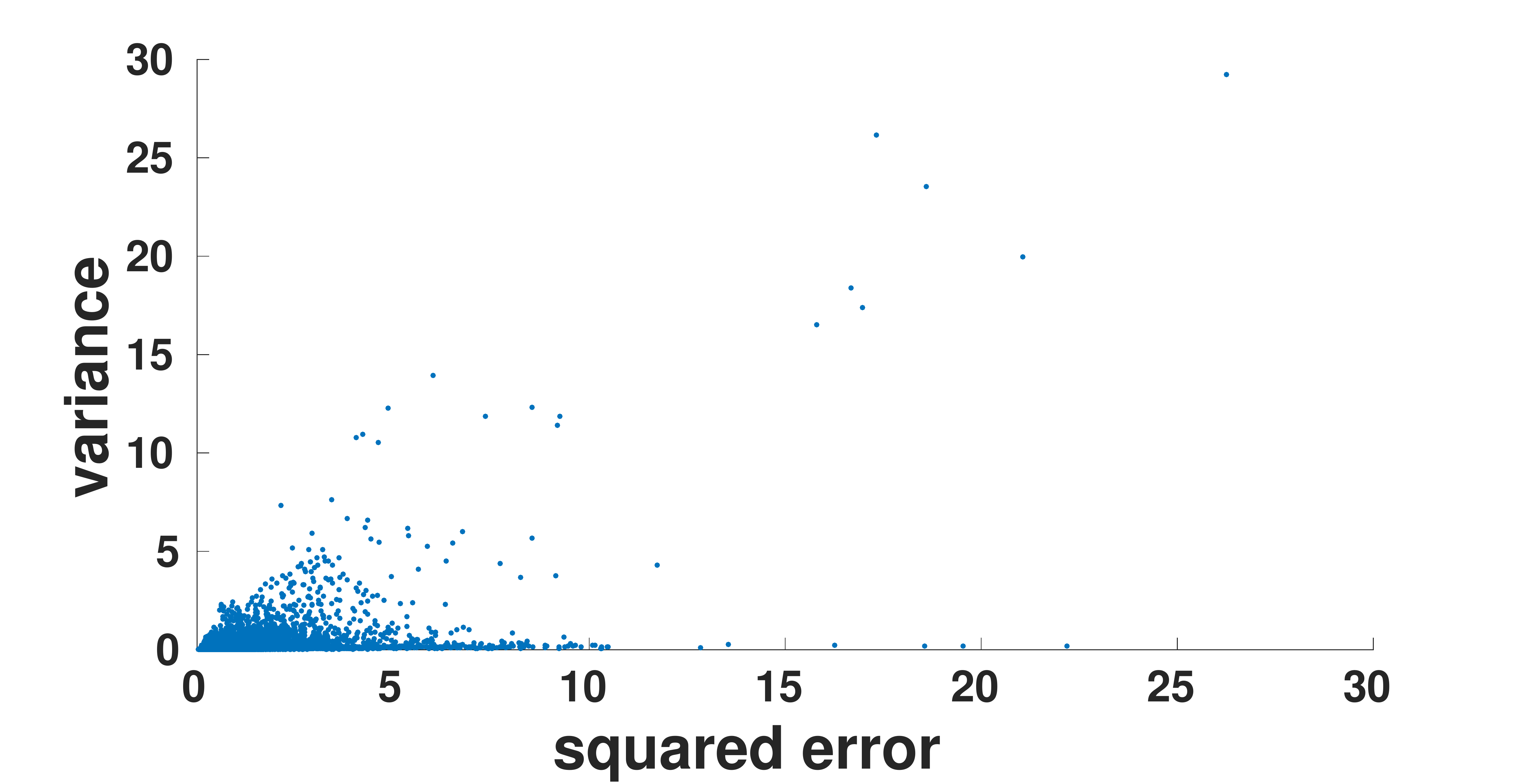}
		&\hspace{-4em}\IncG[width=.2\textwidth,height=.12\textwidth]{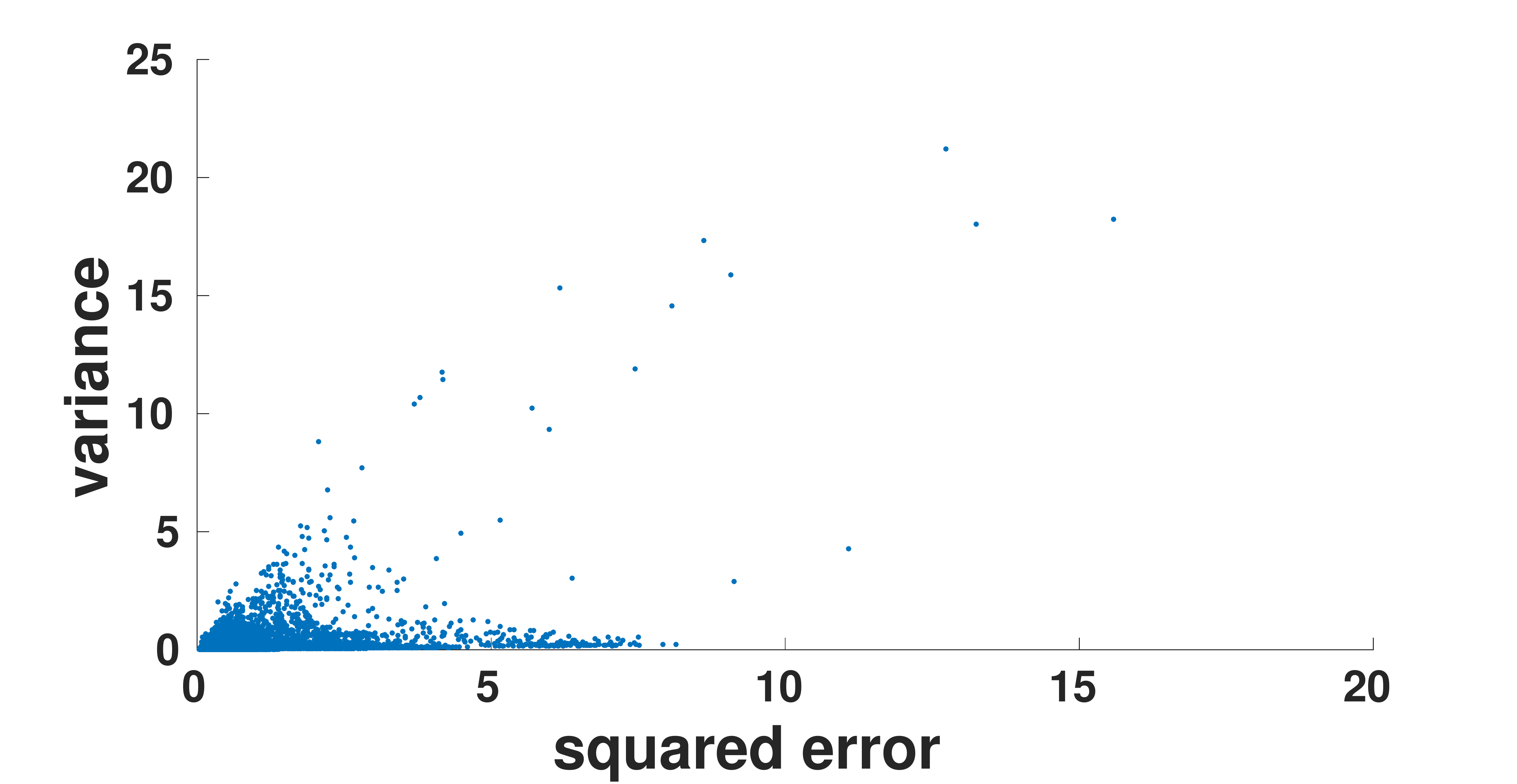}\\
	\end{tabular}
}
	\caption{Correlation plot between the predictive variance and the squared error of the estimated output frames for all the five MC models for the case of speech corrupted with white noise as input}
	\label{fig4}
\end{figure*}

Figure \ref{fig5} shows the variation of SSE with the predictive variance threshold $\mu$, for test data corrupted with all the five seen and three unseen noises for -10 dB SNR. It can be seen that as threshold increases, the performance on unseen noises degrades, while that on seen noises improves. Thus, the threshold $\mu$ can be used to trade-off between the performance of seen and unseen noise cases.

We also evaluated the performance of our algorithms by mixing two unseen noises factory 1 and pink and corrupting the speech file with this new noise at SNRs varying from -10 dB to 10 dB. In another experiment we divided a given speech waveform into three segments and added white, factory2 and factory 1 noise at each segment. Table \ref{table4} shows the performance evaluation of these two experiments. It can be observed that $\mu$-MC algorithm gives superior or comparable performance to Class-conv and Class-MC in all the cases. The algorithm Var-MC and $\mu$-MC algorithm gives superior performance for those cases for which the DNN is less adapted and hence where the correlation between squared error and variance is stronger.
 \begin{table*}[]
 \caption{Performance evaluation of Var-MC and $\mu$-MC algorithms.}
 \resizebox{0.99\linewidth}{2.9cm}{	\begin{tabular}{|l|l|l|l|l|l|l|l|l|l|l|l|}
 		\hline
 		&                                     & \multicolumn{5}{l|}{\textbf{Additive noise factory1+pink}}                                                                                                                                                    & \multicolumn{5}{l|}{\textbf{White-Factory2-Factory1 noises added segment-wise}}                                                                                                                               \\ \hline
 		\textbf{SNR}                     & \textbf{Metric}                     & \textbf{\begin{tabular}[c]{@{}l@{}}Noisy\\ input\end{tabular}} & \textbf{Class-conv} & \textbf{Class-MC} & \textbf{Var-MC} & \textbf{\begin{tabular}[c]{@{}l@{}}$\mu$-MC\\$\mu=0.16$\end{tabular}} & \textbf{\begin{tabular}[c]{@{}l@{}}Noisy\\ input\end{tabular}} & \textbf{Class-conv} & \textbf{Class-MC} & \textbf{Var-MC} & \textbf{\begin{tabular}[c]{@{}l@{}}$\mu$-MC\\$\mu=0.16$\end{tabular}} \\ \hline
 		\multirow{2}{*}{\textbf{-10 dB}} & \textbf{SSE x10\textasciicircum{}4} & 3.77                                                           & 1.06                & 1.03              & \textbf{0.678}  & \textbf{0.892}                                                                   & 4.0                                                            & 0.325               & 0.319             & \textbf{0.309}  & \textbf{0.294}                                                                   \\ \cline{2-12} 
 		& \textbf{SSNR}                       & -8.8                                                           & -6.8                & -6.8              & \textbf{-5.5}   & \textbf{-6.5}                                                                    & -7.2                                                           & -3.5                & -3.5              & \textbf{-2.8}   & \textbf{-3.4}                                                                    \\ \hline
 		\multirow{2}{*}{\textbf{-5 dB}}  & \textbf{SSE x10\textasciicircum{}4} & 1.15                                                           & 0.291               & 0.289             & \textbf{0.244}  & \textbf{0.266}                                                                   & 1.23                                                           & 0.0975              & 0.0963            & 0.159           & \textbf{0.0920}                                                                  \\ \cline{2-12} 
 		& \textbf{SSNR}                       & -7.0                                                           & -4.3                & -4.3              & \textbf{-3.5}   & \textbf{-4.1}                                                                    & -5.2                                                           & -0.9                & -0.9              & -1.0            & \textbf{-0.8}                                                                    \\ \hline
 		\multirow{2}{*}{\textbf{0 dB}}   & \textbf{SSE x10\textasciicircum{}3} & 3.49                                                           & 0.843               & 0.840             & 0.873           & 0.870                                                                            & 3.82                                                           & 0.318               & 0.317             & 0.939           & 0.326                                                                            \\ \cline{2-12} 
 		& \textbf{SSNR}                       & -4.5                                                           & -1.3                & -1.3              & \textbf{-1.0}            & -1.3                                                                             & -2.5                                                           & 1.9                 & 1.9               & 1.2             & 1.9                                                                              \\ \hline
 		\multirow{2}{*}{\textbf{5 dB}}   & \textbf{SSE x10\textasciicircum{}3} & 1.05                                                           & 0.273               & 0.273             & 0.302           & 0.290                                                                            & 1.17                                                           & 0.127               & 0.128             & 0.416           & 0.138                                                                            \\ \cline{2-12} 
 		& \textbf{SSNR}                       & -1.3                                                           & 1.8                 & 1.8               & 1.8             & 1.8                                                                              & 0.7                                                            & 4.5                 & 4.5               & 3.8             & 4.5                                                                              \\ \hline
 		\multirow{2}{*}{\textbf{10 dB}}  & \textbf{SSE x10\textasciicircum{}2} & 3.17                                                           & 1.10                & 1.13              & 1.27            & 1.16                                                                             & 3.58                                                           & 0.71                & 0.72              & 1.67            & 0.75                                                                             \\ \cline{2-12} 
 		& \textbf{SSNR}                       & 2.2                                                            & 4.8                 & 4.8               & 4.6             & 4.8                                                                              & 4.4                                                            & 6.8                 & 6.8               & 6.3             & \textbf{6.9}                                                                              \\ \hline
 	\end{tabular}
 }
 \label{table4}
 \end{table*}
        




\section{Conclusion }
In this work, we propose techniques that use dropout as a Bayesian estimator to improve the generalizability of DNN based speech enhancement algorithms.  The first method uses the empirical mean of multiple stochastic passes through a DNN-MC dropout model trained on multiple noises to obtain the enhanced output. Our experiments show that this technique results in a better enhancement performance, especially on unseen noise and SNR conditions. The second method looks at the potential application of the model uncertainty as an estimate of squared error (SE), for frame-wise selection of one out of multiple DNN models. We devise a method based on a threshold $\mu$ for the predictive variance (Var)
 to switch between a classifier based model selection and predictive variance based model selection. We find that this method gives better enhancement performance compared to classifier based model selection method for unseen noises.
 The main purpose of this work is to see the effectiveness of MC dropout over standard dropout models and hence could be implemented on any state of the art system employing dropout.

\ifCLASSOPTIONcaptionsoff
  \newpage
\fi



\bibliographystyle{IEEEtran}
%
\bibliography{asd2_ref.bib}

\end{document}